%% file: manuscript.tex
\documentclass[%
reprint,
superscriptaddress,
amsmath,amssymb,
longbibliography,
aps,
]{revtex4-1}
\usepackage{graphicx}
\usepackage{verbatim}
\usepackage[version=4]{mhchem}
\usepackage[dvipsnames]{xcolor}
\usepackage{physics}
\usepackage[normalem]{ulem}
\usepackage{pdfpages}
\usepackage{pgffor}
\usepackage{multirow}
\usepackage{bm}
\usepackage{booktabs}
\usepackage{xr}
\usepackage{xr-hyper}
\usepackage{hyperref}
\usepackage{color}

\makeatletter
\AtBeginDocument{\let\LS@rot\@undefined}
\makeatother
\usepackage{caption}
\usepackage{subcaption}
\input{defs}
\newcommand{\rev}[1]{#1}

\usepackage{subfiles}
\begin{document}
\title{Machine learning electronic structure and atomistic properties from the external potential}
\author{Jigyasa Nigam}
\email{jnigam@mit.edu}
\affiliation{Research Laboratory of Electronics, Massachusetts Institute of Technology, Cambridge, 02139 MA, United States of America}

\author{Tess Smidt}
\affiliation{Research Laboratory of Electronics, Massachusetts Institute of Technology, Cambridge, 02139 MA, United States of America}

\author{Genevi\`eve Dusson}
\affiliation{Universit\'e Marie et Louis Pasteur, CNRS, LmB (UMR 6623), F-25000 Besan\c{c}on, France}

\date{\today}
\begin{abstract}
\noindent
Electronic structure calculations remain a major bottleneck in atomistic simulations and, not surprisingly, have attracted significant attention in machine learning (ML). Most existing approaches learn a direct map from molecular geometries, typically represented as graphs or encoded local environments, to molecular properties or use ML as a surrogate for electronic structure theory by targeting quantities such as Fock or density matrices expressed in an atomic orbital (AO) basis. 

\noindent
Inspired by the Hohenberg–Kohn theorem, in this work, we propose an operator-centric framework in which the external (nuclear) potential, expressed in an AO basis, serves as the model input. From this operator, we construct hierarchical, body-ordered representations of atomic configurations that closely mirror the principles underlying several popular atom-centered descriptors. At the same time, the matrix-valued nature of the external potential provides a natural connection to equivariant message-passing neural networks. In particular, we show that successive products of the external potential provide a scalable route to equivariant message passing and enable an efficient description of nonlocal effects. 
We demonstrate that this approach can be used to model molecular properties, such as energies and dipole moments, from the external potential, or to learn effective operator-to-operator maps, including mappings to the Fock matrix from which multiple molecular observables can be simultaneously derived. 
\end{abstract}
\maketitle

\section{Introduction}
Machine learning (ML) is reshaping the modeling of complex physical systems, enabling faster and more accurate simulations and providing new approaches to predicting quantum mechanical properties.
In atomistic modeling, in particular, this progress spans the development of accurate interatomic potentials~\cite{behl-parr07prl,gap2013, nequip, mace,rupp+12prl, shap16mms, drau19prb, behler2021four, musaelian2023learning}, surrogate models for individual properties such as dipole moments~\cite{veit+20jcp, sun2022molecular} or excitation energies~\cite{west-maur21cs, cignoni2024electronic}, and more recently, models that learn elements of electronic structure directly. By targeting the fundamental ingredients of quantum mechanics, such as self-consistent electronic densities, effective single-particle Hamiltonians, or density matrices, ML models provide a route to unified prediction/optimization of multiple observables within a single framework, in addition to accelerating the underlying electronic structure calculations.

Within these workflows, ML can interact with electronic structure in multiple ways. 
Some approaches bypass traditional QM calculations by directly predicting electronic density coefficients~\cite{grisafi2018transferable, rackers2023recipe,koker2024higher}, Fock matrices ~\cite{hegde2017machine, schu+19nc, niga+22jcp, unke2021se3equivariant, zhang2022equivariant, yin2024harmonizingdeephe3, zhong2024universal, haldar2025gears,choudhary2025slakonet}, density matrices~\cite{febrer2024graph2mat, zhang2025dm,li2024image}, or density functionals~\cite{kirkpatrick2021pushing, luise2025skala}. Other strategies integrate ML into a modular hybrid QM-ML pipeline, where electronic operators serve as intermediate optimizable representations, rather than final targets~\cite{cignoni2024electronic, suman2025exploring, tang2024approaching, friede2024dxtb}.  
Despite these conceptual differences, atomic configurations are commonly represented as graphs or point clouds across these workflows, whether the target is a molecular property, such as energy or dipole moment, or a converged electronic operator, such as a Fock or density matrix. While these approaches have been demonstrably successful, they face the challenge of learning complex nonlinear correlations between inputs and outputs, especially in the latter case, where the outputs are high-dimensional matrix representations of quantum operators on a basis, typically chosen to comprise atom-centered orbitals. \rev{Complementary approaches have explored replacing or augmenting purely geometric inputs with electronic quantities, such as effective one-electron reduced density, Fock, exchange matrices, and external potential matrices to predict a wide range of molecular properties. 
For example, Refs.~\citenum{brockherde2017bypassing, shao2023machine} investigate learning the electronic density matrices directly from the external potential, while Refs.~\citenum{fabrizio2021spa,qiao2020orbnet, welborn2018transferability, venturella2025unified} explore the use of 1-RDMs and Fock matrices, or related quantum-mechanical operators as inputs to machine learning models for predicting molecular properties.}
 
Inspired by the Hohenberg–Kohn (HK) theorems~\cite{hohenberg1964inhomogeneous}, which establish that the external potential uniquely determines ground-state electronic quantities and atomistic properties, we reformulate the learning problem by considering the external (nuclear attraction) potential, $\hat{V}_\text{ext}$, as the fundamental input. Rather than operating on the continuous (real-space) potential field, we consider its finite-basis AO discretization, $\mbf{V}_\text{ext}$ (which we denote as $\vext$ for ease of notation), and learn the map from $\vext$ to atomistic properties and self-consistent electronic operators.
We show that in this discretized form, $\vext$ is structurally similar to several commonly used geometric descriptors, allowing a hierarchy of body-ordered features to be derived from it through procedures underlying the construction of atom-centered density descriptors~\cite{musi+21cr} and equivariant message-passing neural networks (MPNNs)~\cite{thom+18arxiv, gilm+17icml}. 

In this work, we show that the external potential can serve as a unified input for learning both atomistic properties and electronic operators. We refer to the model types as Operator-to-Property (\emph{Op2Prop}) and Operator-to-Operator (\emph{Op2Op}) respectively.  For the latter category of models, $\vext$ naturally shares structural properties with the target operators, including rotational symmetry and algebraic constraints such as Hermiticity.
\rev{As previously mentioned, several works ~\cite{brockherde2017bypassing, shao2023machine, bai2022machine} have} tackled a similar problem of mapping the external potential matrices to the density matrix, using kernel-based models, and a recent work~\cite{rana2025learning} even examines the role of kernel choice in model accuracy and transferability. While these models achieve remarkable accuracy in small-data regimes as they are formulated as kernels of finite discretizations of the operators, they are restricted to systems of similar stoichiometry and do not explicitly enforce the algebraic relations and constraints that characterize the targets, such as rotation and permutation symmetries, Hermiticity, etc. 

Implicit in the approach of using $\vext$ as an input is the electronic resolution at which the problem is posed. In the simplest case, and as is routinely done~\cite{shao2023machine}, the input $\vext$ is expressed in the same AO basis as the one used to compute reference properties, or to express target electronic operators. Generally, however, the AO basis used to discretize the input external potential matrix need not match the basis in which the target electronic operators are expressed, and instead can be treated as a tunable model hyperparameter.

In addition to decoupling the resolutions of the input and output bases representations, there is flexibility in how the model is supervised. When tasked with predicting electronic operators (i.e. in the \emph{Op2Op} case), the model can either optimize its matrix elements in a specified basis, or infer an \emph{effective} operator that reproduces a prescribed set of reference observables, potentially obtained from a higher level of theory or more accurate, larger basis-set calculations~\cite{west-maur21cs, cignoni2024electronic, tang2024approaching}. In the latter case (which we term as \emph{effective Op2Op}), the model is constrained only through physical quantities, such as eigenvalues obtained by solving the generalized eigenvalue problem, or expectation values of selected operators, rather than the reference matrix elements themselves.

In the following, we first develop the theoretical framework, showing how the matrix representation of the external potential operator connects to popular geometric descriptors and how matrix operations can implement equivariant message-passing. We then present examples illustrating that $\vext$ matches or exceeds the performance of established descriptors for property prediction and can overcome their limitation in describing long-range interactions. Finally, we use $\vext$ to learn electronic operators (Fock and density matrices), and examine how basis resolution, model architecture, and supervision strategies influence learnability. 

\begin{figure*}[htb]
    \centering
    \includegraphics[width=0.8\linewidth]{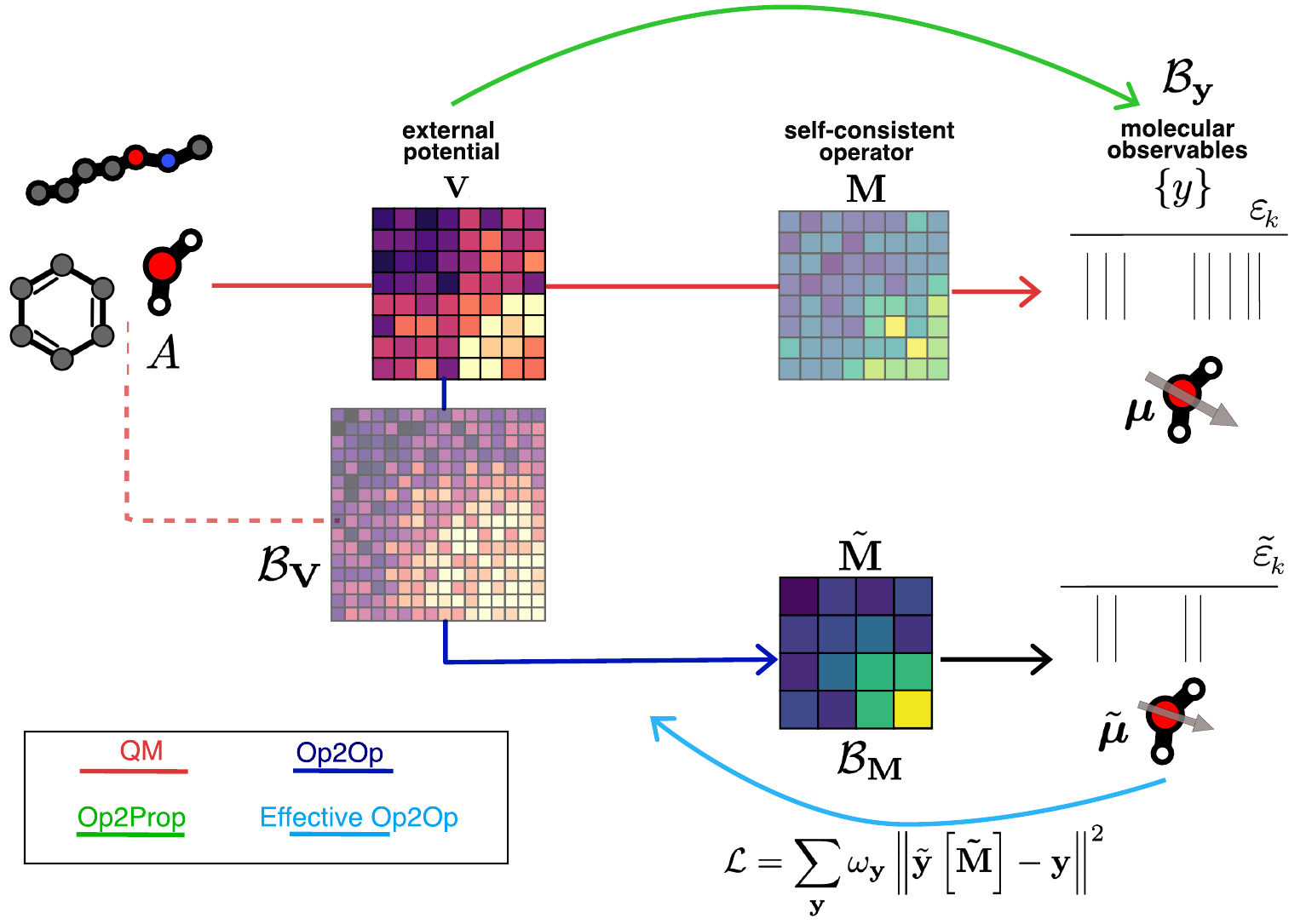}
    \caption{Schematic overview of different models considered in this work. For each molecular configuration, $A$, we aim to predict target properties whose reference values are computed using a quantum mechanical (QM) calculation (top red) in an AO basis set $\mathcal{B}_\mbf{y}$. This calculation produces underlying Fock and density matrices (collectively denoted as $\mbf{M}$), as well as $\vext$ in the top row. $\vext$ can serve as a physically meaningful, naturally symmetry-adapted representation from which target properties may be predicted \emph{Op2Prop} (green). Unlike these property-specific models, \emph{Op2Op} models (dark blue) can map $\vext$ to $\mbf{M}$ in the target $\mathcal{B}_\mbf{y}$ basis, from which properties of interest may be derived through simple analytic operations. Instead of restricting $\vext$ to the basis of reference calculation, we treat it as a tunable hyperparameter and instead use as input, $\vext$ computed in a richer basis set $\mathcal{B}_\mbf{V}$ (red dashed). 
    Similarly, instead of supervising the AO matrices underlying the calculation directly, one can predict a compressed effective representation of the operator on a (usually smaller) basis set $\mcal{B}_\mbf{M}$, which learns an effective projection of the QM calculation on a reduced basis (cyan). 
    }
    \label{fig:op2op_schema}
\end{figure*}

\section{Theory and methods}
For a given atomic configuration, $A$, 
specified by the positions $\{\bR_i\}$ and chemical identities of its atoms $\{Z_i\}$, the nuclear attraction potential operator, $\hat{V}$ is defined as,
\begin{equation}
\label{eq:vext-definition}
\hat{V}(A) = \sum_{a \in A}  \frac{-Z_a}{\left|\mathbf{r}-\mathbf{R}_a\right|},
\end{equation}
where $a$ enumerates the atoms in the configuration with corresponding atomic number $Z_a$. Throughout this work, we restrict our attention to this special case of the external potential. 
The associated matrix representation of the operator when projected in an atomic orbital (AO) basis will be denoted by $\vext$.
In the following, we consider AOs, usually expressed as a product of radial ($n$) and angular functions where real-valued spherical harmonics ($lm$) are used for the latter, and denote the triplet indices $(nlm)$ as $\alpha$.  
When AOs $\chi_\alpha, \chi_\beta$ are centered on atoms $i$, $j$ of structure $A$, the corresponding matrix element of $\vext$ is denoted by $\vext^{\alpha \beta}(A_{i j})$,
and evaluates to 
\begin{equation}
\label{eq:vext-matrix-elements}
    \vext^{\alpha \beta}(A_{ij})= \sum_{a \in A} -Z_a \int \frac{\chi_\alpha^*(\br - \mbf{R}_{i}) \chi_\beta(\br -\mbf{R}_{j}  )}{\left|\mathbf{r}-\mathbf{R}_a\right|} d \mathbf{r}.
\end{equation}
Note that these matrix elements correspond to the projection of $\hat{V}$ for a structure $A$ on basis functions $\alpha$ and $\beta$ respectively centered on atoms $i$ and $j$, that is $\vext^{(i \alpha) (j \beta)}(A)$, which, for ease of readability, we have denoted by $\vext^{\alpha \beta}(A_{ij})$.

For a fixed particle number and a nondegenerate ground state, the first HK theorem (HK1) guarantees that the external potential uniquely determines the many-body ground state wavefunction, and therefore all ground state observables, even though this dependence is not known in closed form. Although the theorem is commonly expressed as a one-to-one correspondence between $\hat{V}$ and the electron density, it implies that both ground state properties and one-particle operators, including the Fock matrix and the one-particle reduced density matrix, can be expressed as functionals of the external potential. In the following, we use the matrix elements of $\hat{V}$, ~\eqref{eq:vext-matrix-elements} as inputs to two complementary modeling strategies that share the same symmetry-adapted external potential inputs but differ in targets, \emph{Op2Prop} models (Fig.~\ref{fig:op2op_schema}, green) that predict molecular properties and \emph{Op2Op} models (Fig.~\ref{fig:op2op_schema}, blue) that learn effective single-particle matrices (Fock or density matrix), treating the input and output basis resolutions as tunable hyperparameters. 
Note that HK1 need not be limited to the nuclear potential we consider in Eq.~\eqref{eq:vext-definition}, but also applies to scalar potentials that are not purely nuclear contributions and may account for additional fields or perturbations. \rev{These potentials can be used as inputs for ML in an analogous manner~\cite{zhang2026transferable}.}

On the other hand, the second HK theorem (HK2) establishes a variational principle for the ground state energy as a functional of the density. While we do not explicitly construct such an energy functional, we invoke its spirit in \emph{effective Op2Op} models (Fig.~\ref{fig:op2op_schema}, cyan), which infer an effective density matrix or Fock matrix that are trained to reproduce reference observables from the underlying ground state QM calculation. \\

\paragraph*{Local descriptors from $\vext$:} Although the nuclear potential is a global operator, and its matrix representation resembles a global atomic descriptor (such as Coulomb matrices), in a localized atomic orbital basis, this representation is sparse and dominated by contributions from spatially local nuclear interactions. Matrix elements decay rapidly with the separation of the atoms on which basis functions are centered (more details in Supplementary Material (SM) section~\ref{label:si-vextlocality}). 
In addition to the sparsity inherited as a consequence of the spatial decay of orbitals, a stricter notion of locality can be explicitly imposed, for example, by setting matrix elements associated with orbital pairs separated by distances $r_{ij}>r_\text{cut}$ to zero if desired. The choice of AO basis set controls the radial and angular resolution,  closely paralleling the construction and tunability of local geometric descriptors such as atom-centered density correlations~\cite{will+19jcp, niga+22jcp2}, the Atomic Cluster Expansion (ACE)~\cite{drau19prb, dusson2022atomic}, 
the Smooth Overlap of Atomic Positions (SOAP)~\cite{bart+13prb}.
Even when targeting specific AO representations of electronic operators or properties resulting from reference calculations performed in another basis, the basis used for $\vext$ can be chosen independently and treated as a tunable hyperparameter, akin to how the radial and angular resolutions of atom-centered geometric features are selected in conventional point cloud ML models. Using a larger basis for the input operator provides additional spatial resolution of the potential, and therefore more expressive features for learning, while remaining computationally inexpensive since $\vext$ requires only non-self-consistent one-electron integral evaluations of Eq.~\ref{eq:vext-matrix-elements}. 

At the same time, the matrix elements of $\vext$ can be understood through the lens of these body-ordered descriptors. Each matrix element $\vext^{\alpha\beta}(A_{ij})$  encodes one-neighbor correlations around the atom pair $(i,j)$, corresponding to a three-body interaction (see SM Section~\ref{sec:body-order-vext} for a detailed derivation).
\\
\paragraph*{Euclidean symmetries of AO representations:} A single-particle operator expressed in an atom-centered basis $\mbf{M}$ transforms under rotations $\Rhat$ of the associated structure as, 
\begin{equation}
\label{eq:matrix-rotation}
    \mbf{M}(\Rhat A) =  D(\Rhat)\mbf{M}(A) D(\Rhat)^\dagger , 
\end{equation}
where $D(\Rhat)$, the Wigner-D matrix, is the appropriate representation of the rotation within the angular-momentum subspace of the atomic-orbital basis.  In particular, the matrix elements associated with each pair of orbitals centered on atoms $i$ and $j$ in structure $A$, characterized by radial quantum numbers ($n_i$,$n_j$), angular momenta  $(l_i, l_j)$, and magnetic quantum numbers $\{ -l_i \leq m_i \leq l_i, -l_j \leq m_j \leq l_j\}$,
\[
\mbf{M}^{(Z_i n_il_i) (Z_j n_j l_j)} _{m_i,  m_j}(A_{ij}) \equiv \rep<i \,n_i l_i m_i | \hat{M}\rep| j \,n_j l_j m_j>,
\]
transform according to the tensor product of Wigner-D matrices,
\begin{multline}
\label{eq:uncoupled-matrix-blocks}
     \rep<i \nlm_i | \Rhat \hat{M} \Rhat^\dagger \rep| j n_j l_j m_j> = \sum_{m_i' = -l_i}^{l_i}  \sum_{m_j' = -l_j}^{l_j}  D^{(l_i)}_{m_i m_i'}(\Rhat) \\ \times   \rep<i nl m_i' | \Rhat \hat{M} \Rhat^\dagger \rep| j n_j l_j m_j'>  D^{\dagger (l_j)}_{m_j' m_j}(\Rhat). 
\end{multline}
This transformation holds when $\mbf{M}$ corresponds to the external potential $\vext$, as well as when it represents the electronic Hamiltonian $\mbf{H}$ or the density matrix $\mbf{P}$. 
We refer to these elements using the shorthand $\mbf{M}^{\mbf{b}}_{m_i m_j}(A_{ij})$ where $\mbf{b}=(Z_i n_i l_i, Z_j n_j l_j)$ labels the chemical species of $i$ and $j$ and the radial and angular quantum numbers. Note that we use the chemical species to label the orbitals and distinguish orbitals with the same $(n,l)$ quantum numbers on different atom types, which are not equivalent. For example, a $1s$ orbital on carbon is functionally distinct from a $1s$ orbital on hydrogen, despite sharing the same quantum numbers.

Following Ref.~\citenum{niga+22jcp}, we transform the product in Eq.~\eqref{eq:uncoupled-matrix-blocks} to irreducible representations (irreps) of rotations indexed by 
$\lambda \in \{ 
|l_i-l_j|, |l_i-l_j|+1, \ldots, l_i+l_j \}, \sigma\in \{-1,1\}, 
\mu \in \{ -\lambda, \lambda+1,\ldots, \lambda\}
$,
\begin{align}
\label{eq:coupled-matrix-block}
    \mbf{M}^{\mbf{b}, \lambda\sigma }_{\mu} (A_{ij}) = \sum_{m_i m_j} \cg{l_i m_i}{l_j m_j}{\glm} \mbf{M}^{\mbf{b}}_{m_i m_j}(A_{ij}) \nonumber \\ \times \delta_{\sigma, (-1)^{l_i +l_j+\lambda} },
\end{align}
where $\cg{\lm}{l'm'}{\glm}$ are the Clebsch-Gordan coefficients, $\sigma$ denotes inversion parity, i.e., whether the resulting irrep behaves as a polar or pseudotensor under spatial inversion (improper rotations).

Furthermore, the matrix elements transform equivariantly under permutations $\boldsymbol{\pi}$ of atom-labels of identical chemical species, as elements corresponding to $\mbf{b}=(Z_i n_i l_i, Z_j n_j l_j)$, $\mbf{M}^{\mbf{b}}_{m_i m_j}(A_{ij})$ are, in general, distinct from $\mbf{M}^{\mbf{b}}_{m_i m_j}(A_{ji})$ (note the exchange of atom labels, while keeping the orbitals fixed) since they depend on the local environments associated with the ordered atom pair. 
A permutation exchanges the atomic indices $i$ and $j$ while keeping the associated orbital channels $\left(Z_i n_i l_i\right)$ and $\left(Z_j n_j l_j\right)$ fixed, but because such relabelings leave the underlying physical structure unchanged, the corresponding matrix elements must be related by symmetry.  We enforce this consistency by constructing symmetric and antisymmetric combinations of Eq.~\eqref{eq:coupled-matrix-block} that transform equivariantly under permutations~\cite{niga+22jcp}. 
$$
\mbf{M}^{\mbf{b}, \lambda\sigma, \eta_{\boldsymbol{\pi}}}_{\mu}(A_{ij}) = \mbf{M}^{\mbf{b},\lambda\sigma}_{\mu}\left(A_{i j}\right)  + \eta_{\boldsymbol{\pi}} \mbf{M}^{\mbf{b},\lambda\sigma}_{\mu}(A_{\boldsymbol{\pi}(i) \boldsymbol{\pi}(j)})
$$
where $\eta_{\boldsymbol{\pi}}= \pm 1$ labels the symmetric and antisymmetric permutation irreps. 
\\
\paragraph*{Algebraic/physical constraints of targets:} 
For \emph{Op2Prop} models, where model targets are molecular properties $y$, such as energies and dipole moments, we consider only symmetry-based constraints. For example, energies are rotational invariants (scalar, $\lambda=0, \sigma=1$), while dipole moments transform as polar vectors ($\lambda=1, \sigma=1$). These symmetries are preserved when the properties are predicted from the corresponding irrep blocks of $\vext$, as explained in the next section. On the other hand, when targeting electronic operators, in addition to basis-dependent rotational symmetries of matrix representations, we must also account for algebraic constraints. For instance, all the electronic matrices considered in this work (including $\vext$) are Hermitian (even symmetric, as the basis functions are real-valued), that is, $\mbf{M} = \mbf{M}^\top$, or in terms of elements
$\mbf{M}^{\mbf{b}}_{m_i m_j}(A_{ij}) = \mbf{M}^{(Z_j n_j l_j, Z_i n_i l_i)}_{m_j m_i}(A_{ji})$ (note the exchange of both atom labels and orbitals, unlike in the previous subsection). 
For the electronic Hamiltonian $\mbf{H}$, Hermiticity ensures a real eigenvalue spectrum obtained through the generalized eigenvalue equation,
\begin{equation}\label{eq:eigv-gen}
    \mbf{H} \mbf{C} = \mbf{S} \mbf{C}\operatorname{diag}\boldsymbol{\varepsilon},
\end{equation}
where each column of $\mbf{C}$ specifies the AO basis coefficients for the eigenvectors (molecular orbitals or MO), $\bm{\varepsilon}$ is the vector of corresponding eigenvalues (MO energies), and $\mbf{S}$ is the matrix describing the overlap of AOs, which are typically non-orthogonal. The eigenvectors $\mbf{C}$, on the other hand, satisfy the condition $\mbf{C}^{\dagger} \mbf{S} \mbf{C}= \boldsymbol{1}$. 

In fact, these quantities underpin all other electronic observables. The single-particle density matrix is first constructed as,
\begin{equation}\label{eq:density-matrix}
    \mbf{P} = \mathbf{C} \operatorname{diag}(\mathbf{f}) \mathbf{C}^\dagger
\end{equation} 
with $\mathbf{f} \in \{0,1\}$ denoting the occupation of each MO. As a consequence of the orthonormality of the MOs, the density matrix satisfies the idempotency relation $\mbf{P} \mbf{S} \mbf{P}=  2\mbf{P}$ (see SM Section \ref{sisec:grassmanian} for the derivation) and lies on a Grassmannian manifold.
Observables corresponding to an operator $\hat{O}$ can then be computed as $\Tr(\mbf{P} \mbf{O})$. 

Within mean-field theories, the electronic structure problem reduces to solving Eq.~\eqref{eq:eigv-gen} self-consistently, as $\mbf{H}$ depends on $\mbf{C}$.
In this work, we focus on predicting the converged solutions ($\mbf{H}$ or $\mbf{P}$), thereby bypassing the iterative self-consistent procedure that constitutes the main computational cost of conventional mean-field methods. This imposes the additional algebraic constraint that the commutator $[\mbf{H}, \mbf{P}]$ be identically zero, which is automatically satisfied when $\mbf{P}$ is constructed following Eq.~\eqref{eq:density-matrix}. 

\par Having described the structure and symmetries of molecular properties and converged mean-field operators, we now turn to the problem of learning these quantities from data.  We use $y$ to denote the target when it is a property output from a converged mean-field calculation, and $\mbf{M}$ to denote the target when it is a converged single-particle matrix (corresponding to either the electronic Hamiltonian $\mbf{H}$ or the density matrix $\mbf{P}$). 
To clearly separate chemical elements, AO basis discretization and the $O(3)$ symmetry indices associated with the inputs and outputs, we rewrite $\mbf{M}^{\mbf{b}, \lambda\sigma }_{\mu} (A_{ij})$ as $\mbf{M}^{\boldsymbol{\gamma}}_{q,  \mu} (A_{ij})$ where $\boldsymbol{\gamma}$ denotes the tuple of chemical species of the pair and the symmetry ($Z_i Z_j, \lambda\sigma$) and $q$ indexes the basis labels $(n_i l_i n_j l_j)$ that take on the role of features, as we will be explained below. 

\subsection{Linear models}
\label{sec:linear-model}
Although the HK theorems establish the ground-state density as a functional of the external potential, and this mapping may, in general, be highly nonlinear, we begin by considering the simplest class of models.

\begin{figure}
    \centering
    \includegraphics[width=1.0\linewidth]{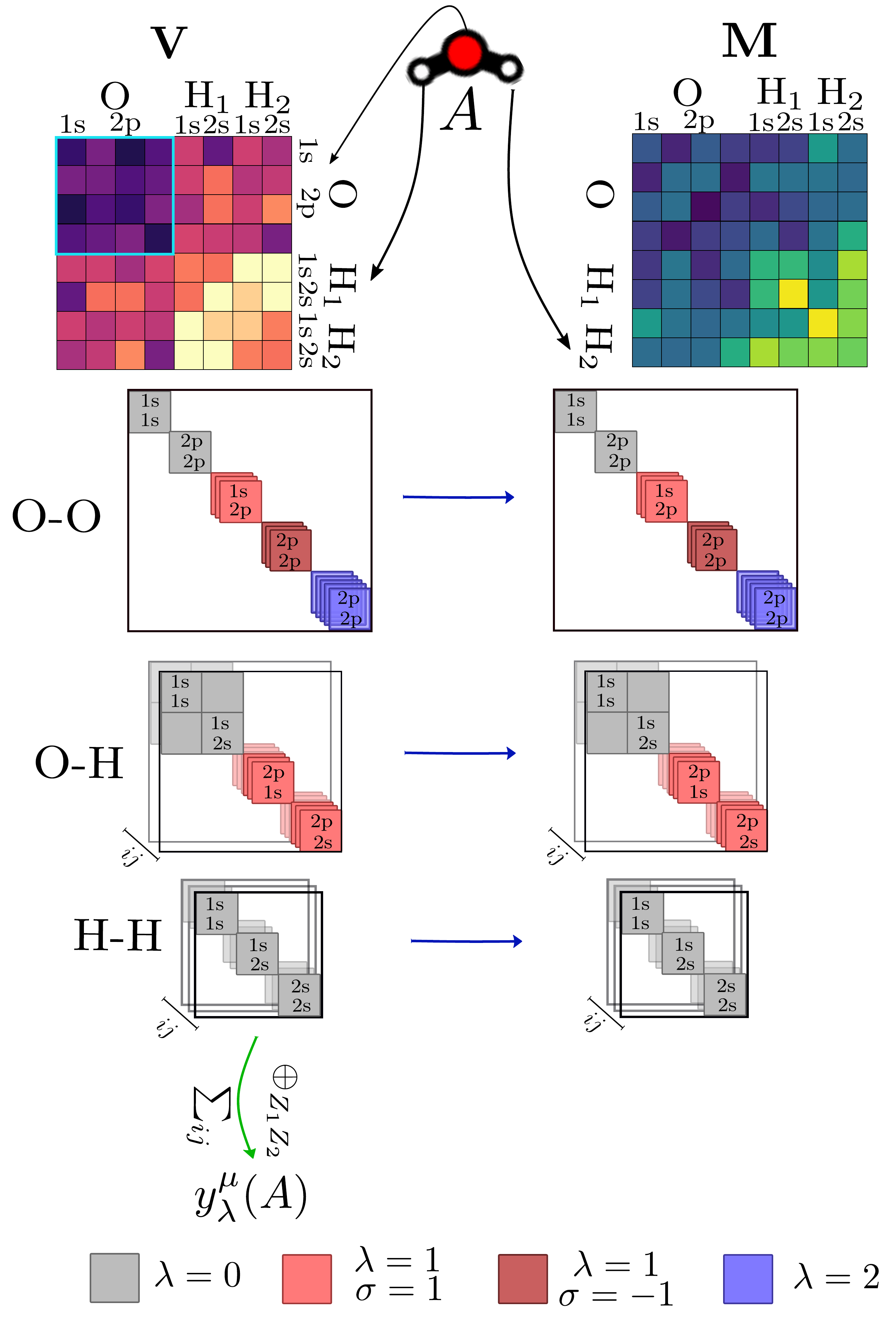}
    \caption{Decomposition of AO matrix representations of atomistic properties and electronic operators \rev{(in a fictitious basis)} into symmetry-adapted blocks. For \emph{Op2Op} models, both the input external potential $\vext$ and output matrices $\mbf{M}$ are decomposed into irreducible representations (irreps) of $O(3)$, indexed by angular momentum $\lambda$ and parity $\sigma$. To ease visualization, we represent $\vext$ and $\mbf{M}$ on the same basis, although this is not necessary for the models presented in this work. For each species pair, the multiplicity of each irrep depends on the orbital pairs that can contribute to the symmetry block, as determined by angular momentum coupling, while the number of such elements in each block is set by the distinct atom pairs $ij$ corresponding to the species. For example, for the species pair (O,H), contributions arise from the atom pairs (O,$\text{H}_1$) and (O,$\text{H}_2$), while for the species pair (H,H), $ij$ corresponds to atom pairs ($\text{H}_1$,$\text{H}_1$), ($\text{H}_1$,$\text{H}_2$) and ($\text{H}_2$,$\text{H}_2$).  A linear model (blue) maps each irrep block of the input to the corresponding irrep block of the output, with block-specific weights.
    For \emph{Op2Prop} models (green), symmetry-adapted blocks of the external potential are summed over all atom pairs and species blocks are concatenated to produce structure-level features for property prediction.
    }
    \label{fig:blockwise-decomposition}
\end{figure}

For \emph{Op2Op} models, as both the input $\vext$ and output matrices ($\mbf{H}$ or $\mbf{P}$) can be decomposed into irreducible representations (irrep) of $O(3)$ as described above, we consider linear models that leverage the shared structure of the inputs and outputs, and map each irrep of the input to the corresponding irrep of the output (Fig.~\ref{fig:blockwise-decomposition}).

Consider, for example, an \ce{H2O} molecule, for which we represent $\vext$ in the def2-SVP basis and the target $\mbf{H}$ in the smaller STO-3G basis. 
The invariant matrix elements associated with orbitals on the oxygen atom, $\boldsymbol{\gamma} = (\text{8},\text{8}, 0,1)$ in $\mbf{H}$ receive contributions from pairs of orbitals $q \in \{ (1s,1s), (1s,2s),(2s,2s),(2p,2p)\}$.The same block ($\boldsymbol{\gamma}= (\text{8},\text{8}, 0,1)$ in the input external potential is significantly richer as basis pairs corresponding to higher angular and radial quantum numbers, such as $(3p,3p), (3d,3d)$ contribute to this block, in addition to those present in the STO-3G basis.

The linear ansatz takes the form,
\begin{align}
\label{eq:linear-model}
    \mbf{M}^{\boldsymbol{\gamma}}_{q, \mu}(A_{ij}) = \mbf{w}^{(\boldsymbol{\gamma})}_{q q'} \mbf{V}^{\boldsymbol{\gamma}}_{ q', \mu}(A_{ij})  ,
\end{align}
where $\mbf{w}^{(\boldsymbol{\gamma})}_{q q'}$ are block specific weights, indicated by the superscript ${(\boldsymbol{\gamma})}$, that act independently within each species and symmetry block. As illustrated in the example above, the basis used to represent $\vext$ is usually larger than that used for the target operator, so the dimensionality of the input feature $q'$ is generally larger than the output index $q$. While these models perfectly preserve the symmetry constraints by construction, they remain limited in expressivity as the linear dependence on blocks of $\vext$ does not capture the global, nonlinear interactions between different species channels or between different atom-pairs, inherent in the self-consistent dependence of $\mbf{M}$ on $\vext$. Nevertheless, linear models are valuable as a conceptual starting point, as they delineate the roles of symmetry, basis resolution, and block structure transparently. In what follows, we build on this construction by incorporating many-body interactions, message passing, and nonlinearities. 

Similarly, in \emph{Op2Prop} models, each symmetry irrep of target property $\mbf{y}^{\lambda\sigma}_\mu(A)$ can be predicted from the corresponding symmetry-adapted block of $\vext$. For a linear model, this takes the form 
\begin{align}
\label{eq:linear-model-property}
    \mbf{y}^{\lambda \sigma}_{\mu}(A) &= \sum_{\boldsymbol{\gamma}} \sum_{ij \in A} \mbf{w}^{(\boldsymbol{\gamma})}_{q} \mbf{V}^{\boldsymbol{\gamma}}_{q, \mu}(A_{ij}) \\
    &= \sum_{\boldsymbol{\gamma}} \mbf{w}^{(\boldsymbol{\gamma})}_{q} 
    \left(\sum_{ij \in A} 
\mbf{V}^{\boldsymbol{\gamma}}_{q, \mu}(A_{ij}) \right),
\end{align}
where the sum over $\boldsymbol{\gamma}$ is restricted to the blocks of $\vext$ corresponding to the target irrep $(\lambda\sigma)$, $A_{ij}$ runs over all atom pairs in the molecule, $q$ indexes the features arising from AO basis used to represent $\vext$, and $\mbf{w}^{(\boldsymbol{\gamma})}_q$ are learnable weights specific to each block. However, as these models are property-specific, they can not be used to derive other observables of interest. 

Noting that \emph{Op2Prop} models can be viewed as a special case of \emph{Op2Op} models by restricting the output irreps and aggregating over appropriate input blocks, we only present equations for \emph{Op2Op} models in what follows.

\subsection{Matrix products and message-passing}
\label{sec:mp-vext}
To approximate the nonlinear dependence of the target on the external potential, we consider truncated polynomial expansions of the form,
\begin{equation}
\label{eq:matrix-product}
    \mbf{M} = f(\vext) \approx w_1 \vext + w_2 \vext^2 + \ldots = \sum_{\kappa = 1}^{\kappa_\text{max}} w_\kappa \vext^\kappa
\end{equation}
where the coefficients are tunable parameters, as is the expansion order $\kappa_\text{max}$, which controls the expressivity of the model. 

The motivation for this choice is twofold. First, the powers of $\vext$ systematically introduce cross-block couplings, since matrix multiplication naturally mixes elements corresponding to different orbital and atom labels, absent in purely linear models~\eqref{eq:linear-model}.  Second, matrix products provide a minimal symmetry-preserving basis to describe these nonlinear effects. Although it may seem counterintuitive, products of matrices in the AO basis transform equivariantly themselves. For instance, consider the transformation of the product of matrices on the same AO basis, $\mbf{M}^2 = \mbf{M} \mbf{M}$.  Using Eq.~\eqref{eq:matrix-rotation}, we obtain 
\begin{equation}
     D(\Rhat)\mbf{M} D(\Rhat)^\dagger D(\Rhat)\mbf{M}D(\Rhat)^\dagger
    =D(\Rhat) {\mbf{M}^2} D(\Rhat)^\dagger. 
\end{equation}
Therefore, $\mbf{M}^\kappa$ undergoes the same transformation as $\mbf{M}$ and thus can be similarly decomposed into irreducible blocks. 
As before, the input $\vext$ and $\mbf{M}$ do not need to share the same AO basis, since in practice we do not map matrices to matrices, but linearly map blocks of $\mbf{M}$ to the symmetry-adapted blocks obtained from each power of the $\vext$, as 
\begin{align}
    \mbf{M}^{\boldsymbol{\gamma}}_{q, \mu}(A_{ij}) &= \sum_{\kappa=1}^{\kappa_\text{max}} \mbf{w}^{(\boldsymbol{\gamma}, \kappa)}_{q q'} (\mbf{V}^\kappa)^{\boldsymbol{\gamma}}_{q' \mu}(A_{ij}),  
\end{align}
which is also equivalent to a single linear map on the concatenation of all the blocks 
\begin{align}
\label{eq:matrix-power-model}
    \mbf{M}^{\boldsymbol{\gamma}}_{q, \mu}(A_{ij}) &= \oplus_\kappa \mbf{w}^{(\boldsymbol{\gamma}, \kappa)}_{q q'} \left(\mbf{V}^{\boldsymbol{\gamma}} \oplus (\mbf{V}^2)^{\boldsymbol{\gamma}} \oplus \ldots \right)_{q' \mu} (A_{ij}).
\end{align}

It must, however, be noted that repeated matrix multiplication can lead to more delocalized representations of the input as information is propagated across multiple orbitals and atomic centers (Fig.~\ref{sifig:receptive-field}) without increasing its dimensionality. This delocalization arises naturally as matrix products implement message passing~\cite{jiang2025demystifying}. Each matrix multiplication introduces a sum over intermediate atomic and orbital indices, thereby propagating information through successive symmetry-adapted matrix blocks. 

For non-equivariant message passing in graph-based models, such as graph-convolutional networks~\cite{kipf2016semi}, it is already well established that repeated multiplications of the adjacency matrix propagate information between connected nodes. Within our framework, $\vext$ plays an analogous role to the adjacency matrix, but with the distinction that it is rotationally equivariant, and thus enables symmetry-aware message passing. Successive matrix powers can be similarly understood as correlations mediated through a chain of atoms over which information is contracted, directly mimicking multi-step message passing on an atomic graph. 
See SM Sections~\ref{sisection:mp}, \ref{sisec:matprod-equiv} for a detailed derivation of equivariance of matrix products and their link to message-passing. \rev{The relationship between matrix products and the message passing has also been explored in matrix function neural networks (MFNNs) introduced in Ref.~\citenum{batatia2023equivariant}. These models construct an intermediate Hermitian, permutation and SO(3)-equivariant matrix operator from learned local node features, apply a learnable matrix function to this operator, and use the diagonal elements of the result to update node features. However, like conventional MPNNs, they take local geometric inputs and predict energies or other molecular properties. MFNNs, in spirit, rely on a mechanism of nonlocal information propagation similar to our work, as the nonlinear operator function can be expressed as an infinite power series expansion (Eq 16 of the Appendix in Ref.~\citenum{batatia2023equivariant}). Therefore, each update step in MFNNs effectively aggregates information from infinitely many matrix powers. In contrast, each matrix product corresponds to a distinct message-passing step in our framework, making the range of nonlocal correlations explicit and systematically controllable (SM Fig.~\ref{sifig:receptive-field}).
}

\subsection{Hierarchical correlation features from AO basis matrices}
\label{sec:tp-vext}
An alternative route to increase the expressive capacity of the input $\vext$ is to generalize matrix products and construct higher-order features using tensor products of symmetry-adapted blocks, analogous to the hierarchical body-order expansions used in atom-centered descriptors or equivariant message-passing neural networks.
Note that the symmetrization of matrix elements from Eq.~\eqref{eq:uncoupled-matrix-blocks}, which correspond to coefficients of an operator expansion in a finite, discrete AO basis, to Eq.~\eqref{eq:coupled-matrix-block}, which represent the symmetry-adapted coefficients of the same expansion, closely parallels the symmetrization of widely adopted atom-density expansions~\cite{niga+20jcp, dusson2022atomic, drau19prb}. We can similarly construct higher-order tensor products to encode many-body correlations at the level of operators. For instance, second-order correlations of $\vext$ can be constructed as,
\begin{equation}
    \label{eq:H-otimes-H}
    (\mbf{V}^{\otimes 2})^{\boldsymbol{\Gamma}}_{q_1 q_2, M} = \sum_{\mu_1 \mu_2} \cg{\lambda_1 \mu_1}{\lambda_2 \mu_2}{\Lambda M} \mbf{V}_{q_1, \mu_1}^{\boldsymbol{\gamma_1}} \mbf{V}_{q_2, \mu_2}^{\boldsymbol{\gamma_2}}, 
\end{equation}
where the label $\boldsymbol{\Gamma} = (Z_1 Z_2 Z_3 Z_4, \Lambda S)$ specifies the species, total angular momentum ($\Lambda$) and parity ($S$) of the block resulting from the combination of blocks of the external potential matrix corresponding to $\boldsymbol\gamma_1 = (Z_1 Z_2, \lambda_1 \sigma_1)$ and 
$\boldsymbol\gamma_2 = (Z_3 Z_4, \lambda_2 \sigma_2)$,
that is $\Lambda \in \{ 
|\lambda_1-\lambda_2|, \ldots, \lambda_1+\lambda_2 \}, S\in \{-1,1\}, 
M \in \{ -\Lambda, \Lambda+1,\ldots, \Lambda\}
$.

Formally, this corresponds to computing the tensor product $\vext\otimes \vext$, thus expanding both along the AO index dimension, as well as the atom indices, as can be seen by expanding the matrix elements on the right-hand side of Eq.~\eqref{eq:H-otimes-H} to include atomic indices $\mbf{V}_{q_1, \mu_1}^{\boldsymbol{\gamma_1}}(A_{i j})$ and $\mbf{V}_{q_2, \mu_2}^{\boldsymbol{\gamma_2}} (A_{i'j'})$. Consequently, the features resulting from  ~\eqref{eq:H-otimes-H} depend on four atomic centers $iji'j'$, in contrast to the two-center nature of the original matrix elements, and may be used to approximate two-electron operators. 

Since the targets considered in this work are one-body operators, we restrict the tensor product to the orbital degrees of freedom, and fix the atomic indices to preserve the two-center structure of the resulting representation, i.e. only consider the contributions from $\mbf{V}_{ q_1, \mu_1}^{(Z_1 Z_2 \lambda_1 \sigma_1)}(A_{ij})$ and $\mbf{V}_{ q_2, \mu_2}^{(Z_1 Z_2 \lambda_2 \sigma_2)} (A_{ij})$.

As with higher-order tensor products of atomic densities, a hierarchy of such operator tensor products may be systematically constructed to form a controllable and systematically improvable set of inputs to model functions of $\vext$. \rev{ The resulting features can then be used as inputs to a linear model following Eq.~\ref{eq:linear-model}. Alternatively, Eq.~\eqref{eq:H-otimes-H} can be implemented within a nonlinear model (such as a message passing neural network), by replacing the traditional geometric inputs with elements of the external potential as described in SM Section~\ref{sisec:mpnn-results}. }

Matrix products correspond to a specific contraction within the full tensor product space, which is obtained by rearranging the coupling in Eq.~\eqref{eq:H-otimes-H}. The tensor product in Eq.~\eqref{eq:H-otimes-H} involves the coupling of the two input irreps $\lambda_1$ and $\lambda_2$ to form the total angular momentum $\Lambda$, through $\cg{\lambda_1 \mu_1}{\lambda_2 \mu_2}{\Lambda M}$. Rather than first forming equivariant blocks and subsequently coupling them, the matrix product first computes rotational invariants associated with the intermediate atomic center and then combines them with the equivariant orbitals on the pair of atoms indexing the matrix element (derivation in the SM Sec.~\ref{sisec:matprod-equiv}).

\rev{\subsection{Nonlinear models}}
\label{sec:nonlinear-scaling}
To introduce simple nonlinearities with matrix products, we apply nonlinear activations to invariants obtained from \emph{all} species and orbital pairs, $\left\{\vext_{q, 0}^{\boldsymbol{\gamma'}}\right\}$,
where $\boldsymbol\gamma'$ enumerates all blocks with $\lambda = 0$ and $\sigma = 1$. 

For each equivariant block with $\lambda > 0$, we multiply the output of this nonlinear function with the corresponding linear equivariant prediction, allowing the invariant features to gate (scale) the equivariant components as,
\begin{equation}
\label{eq:nonlinear-gating}
    \mbf{M}^{\boldsymbol{\gamma}}_{q, \mu}(A_{ij}) = g^{(\boldsymbol{\gamma})} \left(\left\{\mbf{V}^{\boldsymbol{\gamma'}}_{q', 0}\right\}\right) 
    \mbf{w}^{(\boldsymbol{\gamma})}_{q q'} \mathbf{V}^{\boldsymbol{\gamma}}_{q', \mu}(A_{ij}),
\end{equation}
where $g$ is a nonlinear function implemented through a multilayer perceptron (MLP) and can depend on the output block type ${\boldsymbol{\gamma}}$. Although it is possible to construct more sophisticated nonlinear coupling schemes, we only demonstrate this simple gating approach and do not particularly optimize the architecture for simplicity. 

\rev{For models based on tensor products of $\vext$, implemented using an equivariant message passing network, we use nonlinear activations from e3nn~\cite{geiger2022e3nn}. These activations follow a principle similar to the gating mechanism described above. The invariant norm of each output irrep is computed, passed through an MLP, and the result is used to rescale the corresponding equivariant feature. However, unlike the global gating function described above, these nonlinearities are applied after each tensor product. 
}

\subsection{Optimizing effective operators}
\label{sec:effective-operators}
Finally, we consider the \emph{effective Op2Op} models that do not directly optimize the matrix elements of the target operator in the basis of reference calculations ($\mcal{B}_\mbf{y}$), but instead learns an effective representation (also termed intermediate, shadow, or down-folded representations in other contexts) in basis $\mcal{B}_\mbf{M}$, from which target observables are recovered.   
Refs.~\citenum{cignoni2024electronic, suman2025exploring, tang2024approaching}, for instance, map symmetry-adapted geometric inputs to a latent effective Hamiltonian $\mbf{H}$ which is optimized to recover the eigenspectrum or derived quantities from reference calculations in a larger basis set or at a higher level of theory.  
Within the \emph{effective Op2Op} framework, these approaches can be interpreted in the spirit of the Hohenberg--Kohn theorems previously discussed. Although HK1 guarantees that the observables are uniquely determined functionals of $\vext$, the matrix representation of the operators that reproduce these observables need not be unique. Effective operators exploit this non-uniqueness by learning matrix representations that reproduce a desired subset of desired observables (e.g., eigenvalues, expectation values of operators) without necessarily corresponding to the density or matching the reference operator.  Here, this learnt functional (representation of $\mbf{H}$ or $\mbf{P}$) is data-driven and not constrained to arise from a universal density functional or from an underlying variational minimization.  

We first obtain an optimal projection of the target operator by downfolding $\mbf{M}$ from the large basis $\mcal{B}_\mbf{y}$ into a minimal effective basis $\mcal{B}_\mbf{M}$, following the analytical projection derived in SM Section~\ref{sisec:basis-projection}. This effective representation reproduces a subset of target observables of the original matrix, and the model is first trained to recover this downfolded representation.
 
Training then proceeds by minimizing a loss defined over a set of target properties $\mathcal{Y}$, such as eigenvalues or other response functions,
\begin{equation}
\label{eq:effective-op2op-loss}
    \mathcal{L}=\sum_{\mbf{y} \in \mathcal{Y}} \omega_\mbf{y} \left\|\tilde{\mbf{y}}\left[\mbf{\tilde{M}} \right]-\mbf{y}\right\|^2, 
\end{equation}
where $\mbf{y}$ is the reference value of the target property, $\tilde{\mbf{M}}$ is the effective matrix output by the model, $\tilde{\mbf{y}}$ is the property evaluated from the predicted matrix, and $\omega_\mbf{y}$  controls the relative contribution of each term in the loss. 

While any of the modeling strategies discussed so far can be used to predict an effective $\mbf{H}$ or $\mbf{P}$, by supervising the derived observables rather than the matrix elements of the predicted operators, we restrict the examples to an effective $\mbf{H}$ as a demonstration.
 
\section{Examples and results}
This section presents illustrative examples of the framework on selected datasets and using simple model architectures to highlight the role of $\vext$, matrix products, and basis resolution in modeling.
Before examining a few examples of our \emph{Op2Prop} and \emph{Op2Op} models, we empirically examine the expressivity and body-orderedness of $\vext$ and compare it to three-body atom-centered descriptors (denoted as $\rho^{\otimes 2}(A_i)$) 
that have been used extensively for predictive modeling of both atomic properties~\cite{deringer2021gaussian} and electronic operators~\cite{zhang2022equivariant, niga+22jcp}.

\subsection{Basis resolution, body-orderedness of the external potential and message-passing}
\begin{figure}[t]
\centering
\subcaptionbox{}{\includegraphics[width=\linewidth]{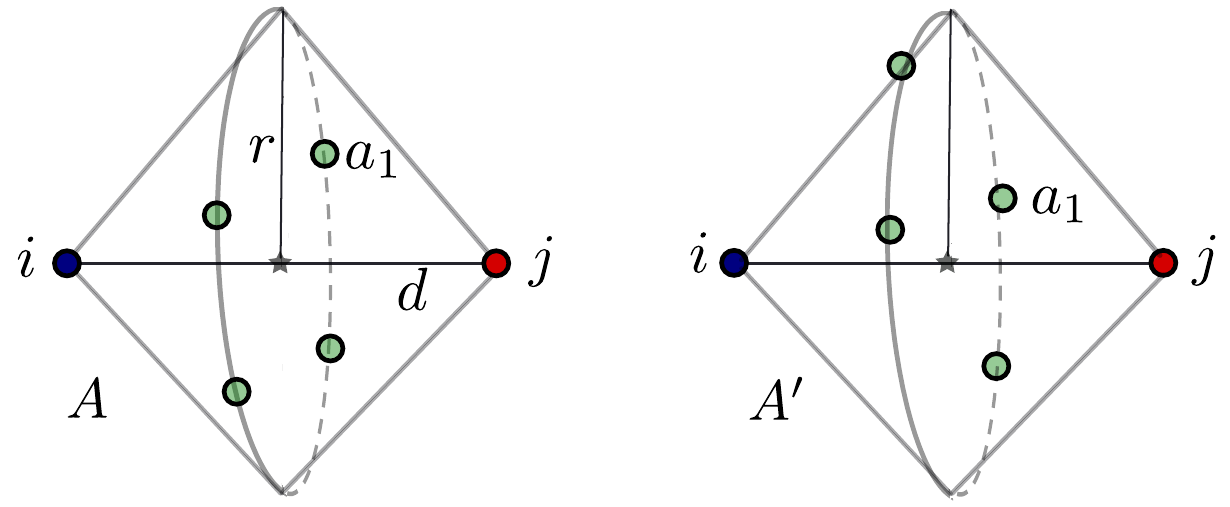}}
\vspace{0.6em}

\subcaptionbox{}{
\begin{tabular}{c|p{3em}|p{3em}|p{3em}}
\hline 
& \multicolumn{3}{c}{Distinguished?} \\
\cline{2-4}
\centering{Configuration} & \centering{$s$} & \centering{$s{+}p$} &  $s{+}p{+}d$ \\
\hline
randomly distorted octahedra & \centering{No} & \centering{Yes} & Yes \\
regular octahedra & \centering{No} & \centering{No} & Yes \\
\hline \hline
\end{tabular}
}
\vspace{1em}
\subcaptionbox{}{\includegraphics[width=\linewidth]{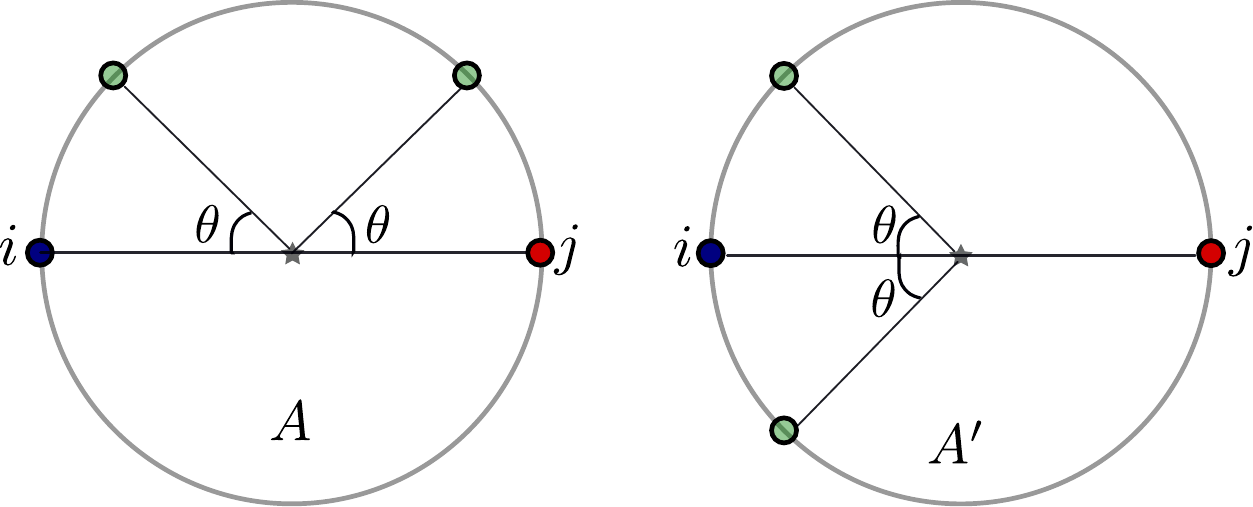}}
\vspace{0.6em}
\subcaptionbox{}{
\begin{tabular}{c|p{3em}|c}
\hline 
$\kappa_{\max}$ & \centering{1} &   2 \\
\hline 
Distinguished?
& \centering No & Yes \\ 
\hline \hline
\end{tabular}
}
\caption{Examples demonstrating the role of basis resolution and matrix products in distinguishing molecular structures. (a) Distorted octahedral geometries, showing a reference atom pair $i, j$ separated by distance $r$ and neighboring atoms placed on a circle in the equatorial plane. (b) Table summarizing whether matrix elements $\vext(A_{ij})$ and $\vext(A'_{ij})$ are distinguishable for basis sets truncated at the listed orbitals for the pair of configurations. Randomly distorted octahedra become distinguishable when $p$ orbitals are included, while regular octahedra require $d$ orbitals due to their higher symmetry. (c) Pair of structures adapted from Ref.~\citenum{pozdnyakov2020completeness} with $\theta = \pi/4$ radians, which have identical three-body correlations and lead to degenerate matrix elements in an $s$-only basis. (d) Table showing that these structures become distinguishable when matrix products ($\kappa_{\max} = 2$) are included, as information from multiple atom pairs is combined through matrix multiplication.
}
\label{fig:degenerate}
\end{figure}

The external potential operator in Eq.~\eqref{eq:vext-definition}, in principle, uniquely specifies the atomic geometry, up to symmetry (rigid rotations, translations, and permutation of identical atoms). Equivalently, in the limit of a complete one-particle basis, the matrix representation in ~\eqref{eq:vext-matrix-elements} contains all geometric information and is itself unique. 
In practice, however, the operator is represented on a finite basis, whose expressive power is limited by its angular and radial resolution. Together with the fact that, in an AO basis, each matrix element of $\vext$ corresponding to the atom pair $A_{ij}$  encodes only one-neighbor correlations relative to the specific pair (SM Section ~\ref{sec:body-order-vext} for a detailed derivation), this finite representation can lead to systematic information loss and to degenerate elements even when pair environments are unrelated by symmetry.
We now consider two complementary examples that disentangle these effects and clarify the role of angular resolution and effective body order in determining the distinguishability of matrix elements.

The first example isolates the effect of basis truncation for a fixed reference pair of atoms. Consider an octahedral structure $A$, consisting of a reference atom pair $A_{ij}$ and a set of neighboring atoms placed on a circle in the equatorial plane perpendicular to the line connecting $i$ and $j$, such that they have the same radial distance relative to the pair, as shown in Fig~\ref{fig:degenerate} a). A second octahedron $A'$ is obtained by randomly varying the angular positions of the atoms on the circle, changing their relative angular arrangement with respect to the pair and mutual radial separation, but preserving their distances to the specified pair. 

In an AO basis set consisting of only $s$ $(l=0)$ orbitals (as is the case for Hydrogen atoms in the STO-3G basis), the matrix element $\vext(A_{ij})$ effectively describes a weighted histogram of functions of pairwise distances of the neighbors relative to the reference pair. Since both distorted octhedra $A$ and $A'$ both have identical radial distributions relative to the pair by construction, the corresponding matrix elements are degenerate, $\vext(A_{ij}) = \vext(A'_{ij})$, despite the fact that the underlying atomic arrangements are geometrically distinct. 
Extending the basis to include $p$ $(l=1)$ orbitals introduces sensitivity to the dipolar components of the neighbor distribution relative to the pair, which lifts this degeneracy. Note that the construction described above need not be restricted to an octahedral arrangement, as the number of atoms on the circle can be varied without affecting the conclusions.

However, for a pair of special configurations, such as when the four neighbors on the circle form a square with inversion symmetry in the plane perpendicular to the pair axis (forming a regular octahedron), 
the $l=1$ contributions vanish by symmetry, causing the degeneracy of elements $\vext(A_{ij})$ and $\vext(A'_{ij})$ to persist even in the larger basis. (In this case, the degeneracy reflects a cylindrical symmetry of the neighbor distribution about the axis joining the pair, rather than the body-ordered resolution insufficiency of the descriptor itself.)  Including $d$ orbitals in the basis set, however, distinguishes the quadrupolar angular moments.  
Generally, while the inclusion of higher angular momentum orbitals improves the sensitivity of the matrix elements to higher-order directional moments of the neighbor distribution, for any finite angular truncation, it is possible to construct an example that would lead to degeneracy. The degeneracy, however, is localized to the chosen reference atom pair; matrix elements associated with other pairs of atoms need not be degenerate.

We now turn to a qualitatively different source of degeneracy, which arises from the three-body (two-center, one-neighbor) nature of the matrix elements and demonstrate how it can be resolved through matrix products, even in the minimal $s$-only basis. Returning to the distorted octhedra in Fig~\ref{fig:degenerate} a) and restricting the basis to only $s$ orbitals,  tensor products of $\vext$ remain functions only of the pairwise distances of neighboring atoms relative to that pair and inherit the same degeneracy as $\vext$. In contrast, matrix powers propagate information through other pairs involving atoms $i$ and $j$, thereby incorporating interatomic distances between neighbors in the equatorial plane that differ between the two structures and resolving the degeneracy.

As a second example, we consider the pair of structures introduced in Ref.~\citenum{pozdnyakov2020completeness}, which are known to be pathologically degenerate under the three-body atom-centered, $\rho^{\otimes 2}(A_i)$ (SOAP) description (Fig.~\ref{fig:degenerate} c), enabling a direct comparison between $\vext$ and SOAP. In these configurations, two atoms lie at diametrically opposite ends of a circle, while the remaining atoms are distributed at different angular positions on the circumference, as shown in Fig.~\ref{fig:degenerate} b), yielding $A$ and $A'$ that are distinct (cannot be mapped onto each other by symmetry) but share identical three-body correlations around the central atom. In this case, we focus on the pair of atoms located at opposite ends of the circle. As before, restricting to the $s$ orbital basis, the matrix element encodes the dependence of the neighbor on their relative distances to the pair center (i.e. the atom at the center). Since these distances are identical for both structures by construction, we obtain matrix elements $\vext(A_{ij})$ equal to $\vext(A'_{ij})$. However, the degeneracy can be lifted without needing to increase the angular resolution, through matrix products, which effectively bring in information from other atom pairs (Fig.~\ref{fig:degenerate} d)). 

\subsection{Datasets and computational details}
In this and the following subsections, we will consider two datasets. The first comprises 1000 distorted water molecules~\cite{gris+18prl, matcloud18c}, which is randomly split into 800 training structures and 200 validation structures. The second dataset we consider is a subset of 1000 molecules, containing C, H, N, and O atoms, from the QM7b dataset~\cite{montavon2013machine}.  

For both datasets, we compute the total energies and dipole moments using PySCF~\cite{sun2018pyscf} at the density functional theory (DFT)/LDA level using the def2-TZVP basis. For each structure, we construct the SOAP descriptor, $\rho^{\otimes 2}(A_i)$ (hyperparameter details in the SM), along with $\vext$ in the cc-PVTZ basis, chosen to produce a comparable basis discretization as the density correlations. Even with this relatively large basis, we obtain roughly six times fewer features from $\vext$ than from SOAP. We drop the explicit structure dependence $(A_i)$ or $(A_{ij})$ from the notation when it is clear from context, to ease readability.

\subsection{Reconstructing representations}
To assess and compare the expressive power of our operator-valued descriptor $\vext$ and 
$\rho^{\otimes 2}$ independently of any particular supervised learning task,  we ask whether one representation can be used to recover the other, and to what extent this reconstruction is possible. We pursue this analysis through a feature reconstruction experiment, similar to Ref.~\citenum{goscinski2021role}. 

To make this comparison as transparent as possible, we restrict the analysis to reconstructions of one representation from another through linear models, as any nonlinearity would effectively introduce infinite body order and obscure the structural difference between them that we aim to probe. Within this linear setting, the reconstruction error effectively conveys how much of one representation lies outside the linear span of the other. Below, we report a normalized reconstruction error $\Delta({\text{X}\to \text{Y}})$, which quantifies the relative error incurred in predicting descriptor $\text{Y}$ from input $\text{X}$ (where $\text{X}, \text{Y}$ in $\{\rho^{\otimes 2}, \vext\}$), 

\begin{align}
\Delta(\text{X} \rightarrow \text{Y}) = 
\frac{\sqrt{\frac{1}{N_\text{test}}\displaystyle\sum_{A\in \text{test}} \left|\tilde{\text{Y}}(\text{X};A) - \text{Y}(A)\right|^2}}{\sigma(\text{Y})},
\end{align}

where $\tilde{\text{Y}}(\text{X};A)$ denotes the optimal linear least-squares reconstruction of $\text{Y}(A)$ from $\text{X}(A)$, and $\sigma(\text{Y})
$ is the standard deviation of the target features over the dataset. Note the asymmetry in $\Delta({\text{X} \rightarrow \text{Y}}) \neq \Delta({\text{Y} \rightarrow \text{X}})$, which reflects the difference in expressivity of the two representations.

As explained in Section~\ref{sec:mp-vext}, computing algebraic powers of $\vext$ corresponds to message-passing and progressively richer many-body information. To verify that this indeed leads to a change in expressive power of $\vext$, we analyze how the reconstruction errors $\Delta({\vext^{\kappa_\text{max}} \to \rho^{\otimes 2}})$ and $\Delta({ \rho^{\otimes 2} \to \vext^{\kappa_\text{max}} })$ vary as additional matrix powers are included. 

\begin{figure}
    \centering
    \includegraphics[width=1.0\linewidth]{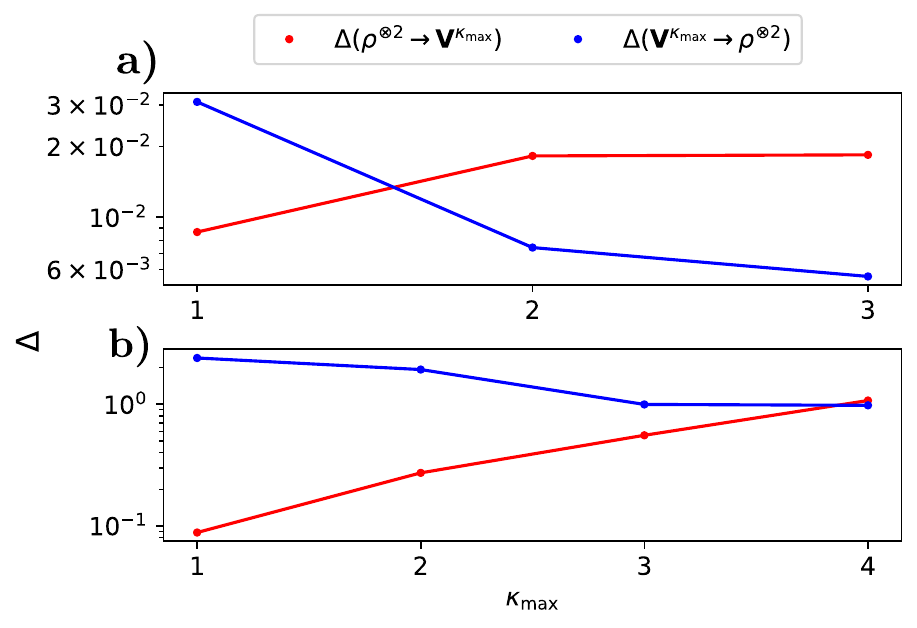}
    \caption{Considering only the invariant features ($\lambda = 0, \sigma=1$), for water molecules (a), organic molecules from the QM7 dataset (b), we plot the error of constructing atom-centered density correlations $\rho^{\otimes 2}$ using input $\vext^{\kappa_\text{max}}$  (blue) and vice-versa (red). The reconstruction error is reported as a function of the matrix power, which makes the representation of the external potential increasingly nonlocal. 
    }
    \label{fig:reconstruction-qm7}
\end{figure}

In Fig.~\ref{fig:reconstruction-qm7}, we see that in the absence of higher matrix powers, $\rho^{\otimes 2}$ reconstructs $\vext$ more accurately than vice-versa (although it should be noted that we recover about six times fewer features using $\vext$  than SOAP due to the basis discretization).
Although both SOAP and $\vext$ both encode three-body information (SM Section~\ref{sisection:acdc-vext}), they do so in markedly different ways. SOAP describes three-body correlations via products of atom-centered neighbor densities, whereas $\vext$ captures weighted three-body interactions mediated by the Coulomb kernel. Moreover, the radial and angular bases in which these two representations are expressed are different, and the descriptors are not explicitly normalized, resulting in features that reside on different numerical scales. 

Despite these differences, $\vext$ becomes increasingly expressive as higher matrix powers are included. As higher powers of $\vext$ are included and concatenated, we observe a systematic decay in $\Delta({\vext^{\kappa_{\max}} \rightarrow \rho^{\otimes 2}})$, indicating that it becomes increasingly easier to construct $\rho^{\otimes 2}$ from $\vext$. At the same time, $\Delta({\rho^{\otimes 2} \rightarrow \vext^{\kappa_{\max}}})$ increases, reflecting the growing structural complexity of $\vext$ as additional nonlocal information is included through message passing. 

\subsection{Predicting molecular properties from external potential}
As discussed in Section~\ref{sec:linear-model}, \emph{Op2Prop} models predict each symmetry irrep of the target property from the corresponding symmetry-adapted block of $\vext$. For energies, which are rotational invariants, we isolate the $\lambda=0, \sigma=1$ blocks from the symmetry-adapted decomposition (Eq.~\eqref{eq:coupled-matrix-block}), concatenate the pairwise species blocks, and sum over all the corresponding atom pairs to produce a structure-level feature vector, $\vext^{\lambda =0, \sigma=1}(A)$. We enrich the representation by concatenating symmetry-adapted blocks from successive powers of the external potential matrix (Section~\ref{sec:mp-vext}), corresponding to features from Eq.~\eqref{eq:matrix-product} truncated at the third power.

Similarly, we obtain a structure-wise SOAP $\rho^{\otimes 2, \lambda=0, \sigma=1}(A)$ feature by summing over all atom-centered contributions. For convenience of notation, we will drop the index names from the superscript, as well as the structure label, and denote these features as $\vext^{\kappa_\text{max}, 01}$ and $\rho^{\otimes 2, 01}$.  
Dipole moments can be analogously modeled, with the exception that they transform as $\lambda =1, \sigma=1$ $O(3)$ irreps, so we accordingly use  $\vext^{\kappa_\text{max}, 11}(A)$, $\rho^{\otimes 2, 11}(A)$ as inputs. 

Using these structural features, we train linear models on the corresponding target properties. Despite its compactness, $\vext$ matches or exceeds the predictive performance of SOAP across both datasets. We report the root-mean-square error (RMSE) computed as
\begin{equation}
    \text{RMSE} = \sqrt{\frac{1}{N_\text{test}} \sum_{A \in \text{test}} \left\|\tilde{\mathbf{y}}(A) - \mathbf{y}(A)\right\|^2},
\end{equation}
where $\tilde{\mathbf{y}}(A)$ and $\mathbf{y}(A)$ denote the predicted and reference property values for structure $A$, respectively, and $N_\text{test}$ is the number of test structures.

For the water dataset, the resulting RMSEs on the validation set are approximately 2.03 meV for total energies and 1.7 mD for dipole moments when using SOAP, and 1.53 meV for energies and 2.1 mD for dipoles when using $\vext$. For the QM7b subset, $\vext$ achieves roughly an order of magnitude lower error than SOAP for energies, 152 meV compared to 2.01 eV, but a similar error of 140 mD on dipole moments.

\begin{figure}
    \centering
    \includegraphics[width=1.0\linewidth]{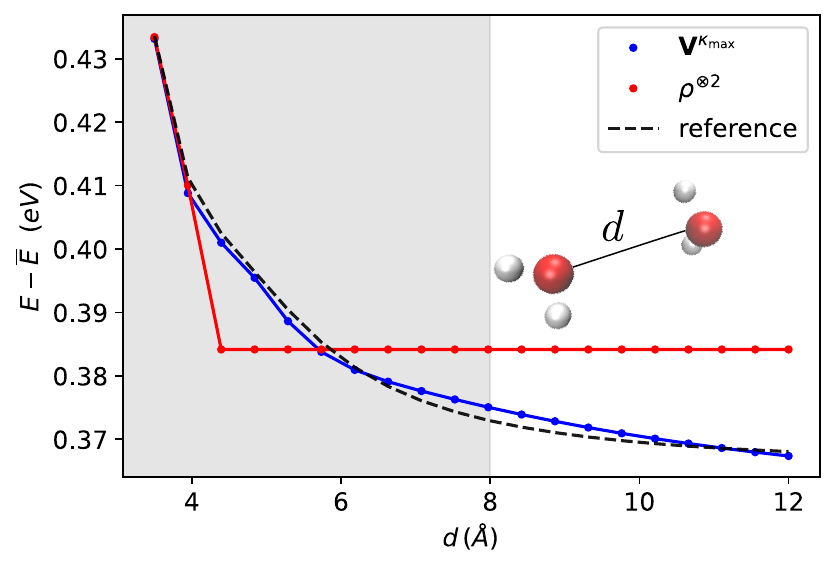}
    \caption{Interaction energy for water dimers (total energy baselined by the mean $\overline{E}$) as a function of the distance $d$ between the two monomers. The shaded region shows distances incorporated in the training set. Reference values (black dashed) compared against predictions from the atom-centered descriptors $\rho^{\otimes 2}$ and products of the external potential matrix up to maximum algebraic power $\kappa_\text{max} = 3$ (blue).
    }
    \label{fig:water-dimer}
\end{figure}

Finally, to demonstrate the effective message passing implemented through matrix products of $\vext$, we construct a dataset of water dimers by selecting 50 random pairs of molecules from the water dataset described above. For each dimer, we generate a one-dimensional trajectory by systematically increasing the intermolecular separation between the two monomers from 3.5~\AA{} to 12~\AA{} in twenty steps, and for each configuration, compute the total energy at the DFT/PBE level using the def2-TZVP basis.
The dataset is partitioned such that configurations with intermolecular separations up to 8~\AA{} are used for training, and the remaining portion of each trajectory is reserved for validation, resulting in 600 dimers in the training set and 400 in the validation set.
Atom-centered descriptors are constructed with a cutoff radius of 4~\AA{}, and the sum over atoms to construct the feature per structure is restricted to only oxygen atoms. As a consequence, when the intermolecular separation exceeds the cutoff distance, the local atomic environments become indistinguishable, and a model based on $\rho^{\otimes 2}$ is unable to capture the long-range decay of the interaction energy (Fig.~\ref{fig:water-dimer}, red). On the other hand, for $\vext$, we construct features from polynomials truncated at the third power (corresponding to symmetry-adapted blocks from $\vext$, $\vext^2$, and $\vext^3$ as described in Section~\ref{sec:mp-vext}), where the sum over atom pairs is restricted such that $i$ indexes only oxygen atoms, while $j$ runs over all allowed neighbors. Even linear models using these inputs are able to reproduce the long-range tail of the interaction energy (Fig.~\ref{fig:water-dimer}, blue). 
 
\par Building on our analysis of $\vext$, its algebraic powers, and their connection to geometric descriptors, as well as its performance in predicting molecular properties, we proceed to the task of learning electronic-structure operators, namely the Fock matrix $\mbf{H}$ and the density matrix $\mbf{P}$, within our \emph{Op2Op} models. 

For the examples in the following subsections, we define the mean square error associated with the overall prediction $\tilde{M}$ of the target matrices over a given test set as 

\begin{equation}
\text{MSE}_\text{\mbf{M}} = \frac{1}{N_{\text{test}}} \sum_{A\in \text{test}} \frac{1}{N_{\mbf{M}(A)}}
 \sum_{i \alpha  j\beta}
\lvert \mbf{M}^{\alpha \beta}(A_{ij}) - \tilde{\mbf{M}}^{\alpha \beta}
(A_{ij})\rvert^2 
\label{eq:mse-full}
\end{equation}

where $N_{\text{test}}$ is the number of structures in the test set, $A$ enumerates the molecular structures in this set, and $N_{\mbf{M}(A)}$ indicates the size of the corresponding output matrix. The MSE on other properties is similarly defined. In several examples below, we report the relative MSE, which is obtained by normalizing the MSE by the standard deviation of the reference values.

\rev{
\subsection{Comparing tensor products and matrix products}
\label{sec:results-compare-with-mpnn}
As a first example of \emph{Op2Op} models, we compare different strategies of using $\vext$ as model inputs, i.e., the tensor product and matrix product formulations described in Sections~\ref{sec:mp-vext} and ~\ref{sec:tp-vext}, against state-of-the-art graph neural networks (GNNs) that operate only on local geometric features. We consider a dataset of 1000 \ce{C14H10} molecules reported in Ref.~\citenum{cignoni2024electronic, matcloud-anthracene}. As anthracene combines extended geometric length scales, with a maximum interatomic separation of 9.73~\AA{} across the dataset, with electron delocalization across three fused benzene rings, it makes for a useful benchmark to assess how different message passing schemes capture both nonlocal geometric and electronic effects. 

The resulting RMSE on predictions of the Fock matrix ($\mbf{H}$) in the STO-3G and def2-SVP basis are summarized in Table~\ref{tab:model_comparison}. Models incorporating $\vext$ consistently outperform geometry-only message passing (denoted by GNN), reducing the RMSE by nearly a factor of two relative to GNN when targeting $\mbf{H}$ in the STO-3G basis. 
\begin{table}[h]
\centering
\begin{tabular}{lccc}
\hline
\hline
Basis set & GNN & GNN$(\vext)$  & $\vext^{\kappa}$ \\
\hline
STO-3G & 1.1 $\times 10^{-2}$ & 6.2 $\times 10^{-3}$  &  4.4 $\times 10^{-3}$\\
def2-SVP & 3.5 $\times 10^{-2}$ &3.1 $\times 10^{-2}$ &  2.9 $\times 10^{-2}$\\
\hline
\hline
\end{tabular}
\caption{\rev{RMSE per matrix element (Hartree) of validation $\mbf{H}$ matrices in the STO-3G and def2-SVP basis. For the latter, targets were projected onto the STO-3G basis following the procedure discussed in SM Section~\ref{sisec:basis-projection}. GNN($\vext$) uses $\vext$ in the STO-3G basis, whereas models based on matrix products compute powers of $\vext$ in the def2-SVP basis up to $\kappa = 4$. Additional details are provided in SM Section~\ref{sisec:mpnn-results}.
}}
\label{tab:model_comparison}
\end{table}

As mentioned above, the two $\vext$ approaches differ in how they build expressive equivariant representations, and therefore, in computational scaling. In GNN($\vext$), the initial irreducible representations of $\vext$ are combined through tensor products during message passing, allowing the network to construct higher-order correlations of the input. A single tensor product generates all pairwise couplings between irreducible representations (c.f. Eq.~\eqref{eq:H-otimes-H}). At the following message passing layer, the model generates couplings of couplings, and so on, leading to a rapidly increasing computational and memory cost. 

Models based on $\vext^{\kappa}$, on the other hand, avoid the recursive  Clebsch-Gordan coupling. They compute products of $\vext$ before decomposing the resulting matrix into irreducible representations  (as described in SM Section~\ref{sisec:matprod-equiv}). As a result, the number of irreps in the model remains fixed irrespective of the matrix power or effective number of message-passing steps.  
This allows models based on the matrix product to be initialized with richer inputs (for instance, through larger basis-set representations of $\hat{V}$), instead of relying on increasingly complex tensor-product operations to construct higher-order equivariant features. 

The results in Table~\ref{tab:model_comparison} indicate that models based on  $\vext^\kappa$ are comparable to, or improve the accuracy of models based on either message passing approach. 
Motivated by this favorable cost-accuracy tradeoff, we focus exclusively on matrix-product models in the following sections.
}

\subsection{Fock and density matrix predictions for water molecules}
For the water dataset described above, we compute the Fock, overlap, and density matrices using DFT/LDA level with the STO-3G and def2-TZVP basis sets using PySCF~\cite{sun2018pyscf}. In addition, the external potential operator $\vext$ is evaluated in multiple atomic orbital bases, including STO-3G, def2-SVP, def2-TZVP, and cc-pVTZ.

\begin{figure}[hb]
    \centering
    \includegraphics[width=0.9\linewidth]{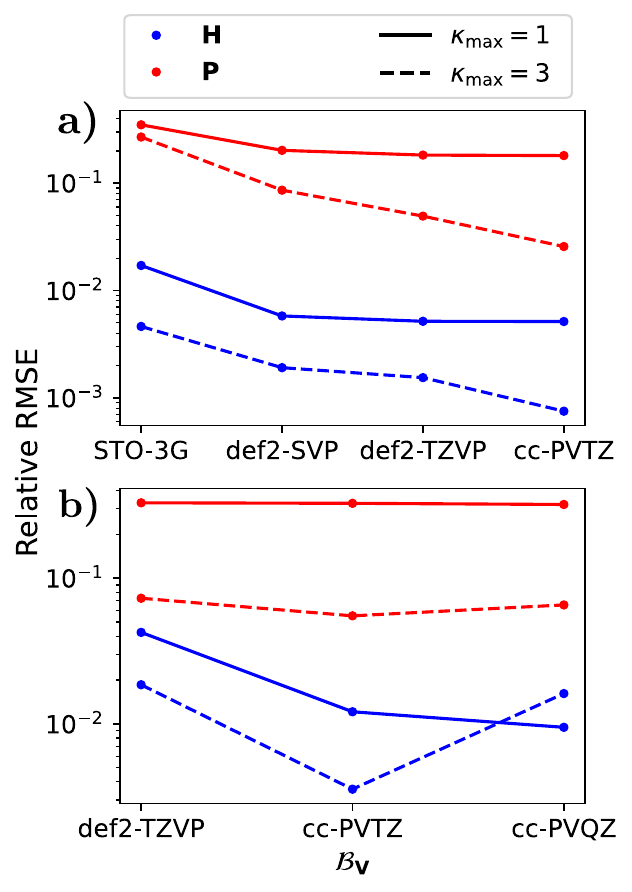}
    \caption{Relative RMSE per matrix element (Eq.~\eqref{eq:mse-full}, normalized by the standard deviation of the target) for the \ce{H2O} dataset as a function of the basis $\mathcal{B}_\mathbf{V}$ used for the input $\vext$ in the linear model Eq.~\eqref{eq:linear-model}. The target operator $\mbf{M}$ (Fock and density) matrices are in STO-3G (a) and def2-TZVP (b) bases, respectively. 
    }
    \label{fig:vext-basis-ablation-water}
\end{figure}

To systematically assess the effect of the input basis resolution on model performance, we train a simple linear model mapping symmetry-adapted blocks of $\vext$ to the target operator matrices and report the resulting RMSEs in Fig.~\ref{fig:vext-basis-ablation-water}.

As expected, the errors on the Fock matrix are generally lower than those on the density matrix, as $\vext$ and $\mbf{H}$ are similarly local operators (~\ref{sifig:qm7sto3g-matrixnorms-distance}), while a linear model built on $\vext$ struggles to capture the nonlocality of $\mbf{P}$ elements. The accuracy is further improved by considering matrix products as inputs (dashed lines in Fig.~\ref{fig:vext-basis-ablation-water} show that incorporating the third power yields an improvement of almost five times with the largest input basis set).

\subsubsection{Op2Op models}
\begin{figure}
    \centering
    \includegraphics[width=1.0\linewidth]{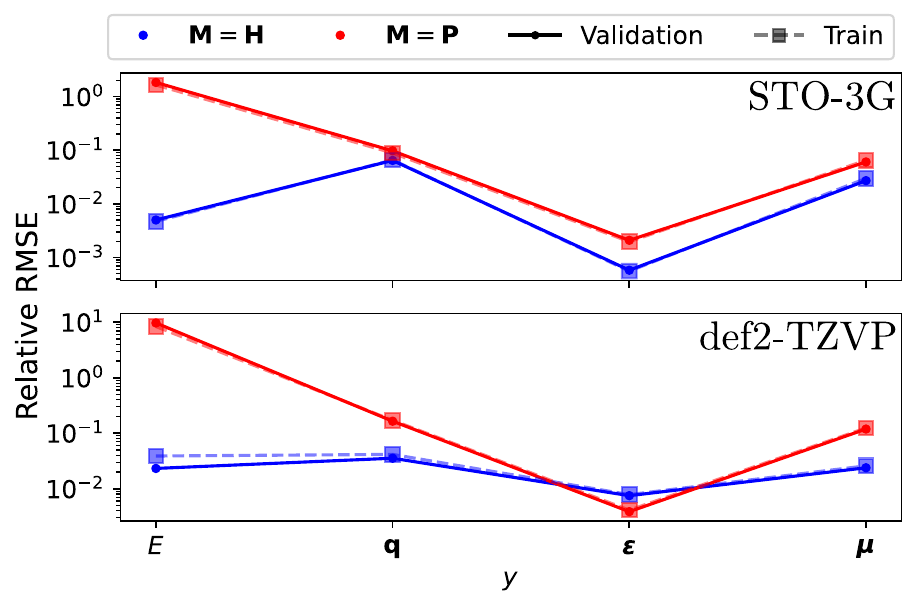}
    \caption{Relative RMSE for the validation set on molecular properties, namely total energy (E), atomic L\"owdin charges $(\mbf{q})$, eigenvalues ($\boldsymbol{\varepsilon}$) and dipole moments ($\boldsymbol{\mu}$) for models trained on the Fock (blue) and density (red) matrices for \ce{H2O} dataset in STO-3G basis (top) and def2-TZVP basis (bottom).
    } 
    \label{fig:vext-direct-water}
\end{figure}

We now consider nonlinear models that incorporate matrix products and nonlinear gating as described in Section~\ref{sec:nonlinear-scaling}. For these models, we use cc-pVTZ as the input basis $\mcal{B}_\mbf{V}$ to express the external potential, and employ the same architectures for models targeting both $\mbf{H}$ and $\mbf{P}$, without attempting to optimize hyperparameters. We supervise the symmetry-adapted matrix elements directly using Eq.~\eqref{eq:mse-full} as the loss in the model. Models trained on the Fock matrix incur an RMSE of 119.7 meV and 8.7$\times 10^{-3}$
respectively on the Fock and the density matrix in the STO-3G basis (Fig.~\ref{fig:vext-direct-water}) and 144.5 meV, 1.4$\times 10^{-2}$ for the def2-TZVP basis. In comparison, models trained on the density matrix, incur an RMSE of 416.6 meV, 3.1$\times 10^{-2}$ on the Fock and density matrix in the STO-3G basis, 1.14 eV and 9.5$\times 10^{-3}$ on the two matrices in the def2-TZVP basis. (Note that the elements of the Fock matrix have the units of energy, but the density matrix is dimensionless).

In addition to errors on the matrix elements themselves, we compute errors on quantities derived from the predicted matrices following diagonalization as in Eq.~\eqref{eq:eigv-gen}, including eigenvalues and expectation values of selected operators.
We report relative errors by normalizing the RMSE by the standard deviation of the target quantities, as shown in Fig.~\ref{fig:vext-direct-water} for molecular properties including total energy ($E$), atomic L\"owdin charges ($\mbf{q}$), eigenvalues ($\mbf{\varepsilon}$), and dipole moments ($\mbf{\mu}$). 
Models trained on the Fock matrix consistently show lower errors in derived properties than models trained directly on the density matrix. 

The predicted matrices are always Hermitian by construction, as we only predict matrix elements in the upper triangular portion and enforce Hermiticity through symmetry. When $\tilde{\mbf{P}}$ is constructed from the predicted $\tilde{\mbf{H}}$ using Eq.~\eqref{eq:density-matrix}, it lies on the Grassmannian manifold and commutes with $\mbf{H}$ by construction. In contrast, when the model is trained to directly predict the density matrix, Grassmannianity is not enforced as a constraint. As a result, the predicted $\tilde{\mbf{P}}$ can deviate from the Grassmannian manifold. We quantify this deviation through the relative violations of idempotency, $\delta_\text{Grassman} = \left\lVert \tilde{\mbf{P}} \mbf{S} \tilde{\mbf{P}} - 2 \tilde{\mbf{P}} \right\rVert /
\left\lVert 2 \tilde{\mbf{P}} \right\rVert$. $\delta_\text{Grassman}$ takes values of 1.4$\times 10^{-2}$  for the STO-3G basis and 0.99 for the def2-TZVP basis, indicating that the predicted density matrix is close to projector-like in the minimal basis but deviates strongly from the Grassmann manifold in the larger basis.

We find that this deviation correlates with the errors in derived properties, and projecting predicted density matrices back onto the Grassmannian manifold, by enforcing idempotency, systematically improves the accuracy of derived properties (SM Sec.~\ref{sisec:water-results}). This indicates that enforcing this constraint may be important for accurate property prediction, as was similarly observed in Ref.~\citenum{zhang2025dm}.  

It is also possible to relax the Hermiticity constraint as a hard inductive bias~\cite{qian2025equivariant}, and it would be interesting to explore its consequences on the resulting properties, but we leave this investigation for future work.

\subsubsection{Effective Op2Op models}
As discussed in Section~\ref{sec:effective-operators}, we can model the Fock operator on an effective basis $\mcal{B}_\mbf{M}$ even though the target properties are computed with a larger basis $\mcal{B}_\mbf{y}$, allowing the choice of basis set representation of $\mbf{M}$ to be treated as a model hyperparameter. To do so requires a differentiable evaluation of the properties from the predicted matrices during training, so that the loss on these derived observables can be propagated back to $\mbf{M}$. Our models interface seamlessly with differentiable electronic structure codes, such as \textsc{PySCFAD}~\cite{zhang2022differentiable}, which can then be used to evaluate molecular properties such as L\"owdin charges, dipole moments from the predicted matrices. 
As a demonstration, we consider learning an effective Fock matrix optimized to reproduce eigenvalues, L\"owdin charges and dipole moments. When projecting from the $\mcal{B}_\mbf{y} = $ def2-TZVP basis to the $\mcal{B}_\mbf{M} = $ STO-3G basis, we can only describe 7 states, corresponding to the size of the matrix in the STO-3G basis. 
These seven
states are obtained through the selection described in SM Section~\ref{sisec:basis-projection}. 

\begin{figure}
    \centering
    \includegraphics[width=1.0\linewidth]{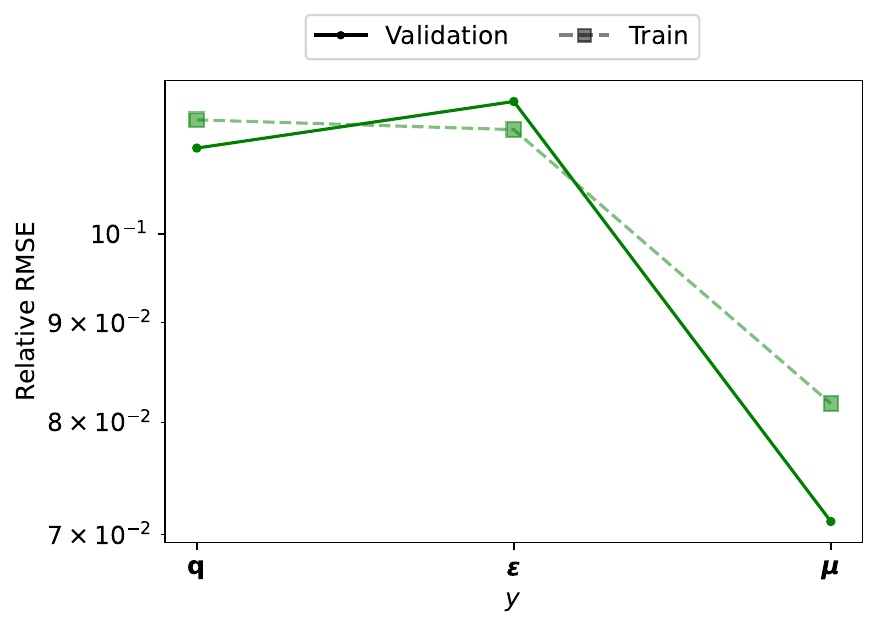}
    \caption{Relative RMSE on L\"owdin charges ($\mbf{q}$), eigenvalues ($\boldsymbol{\varepsilon}$) and dipole moments ($\boldsymbol{\mu}$) from the \emph{effective Op2Op} model. The reference observable values were computed using the $\mcal{B}_\mbf{y}=$ def2-TZVP basis set, while the model predicted the downfolded Fock matrix in the $\mcal{B}_\mbf{M}=$ STO-3G basis. The input $\vext$ in the cc-PVTZ basis, with matrix powers ~\eqref{eq:matrix-product} truncated at $\kappa_\text{max}=2$. The RMSEs have been normalized by the standard deviation of each observable. } 
    \label{fig:vext-indirect-water}
\end{figure}

Figure~\ref{fig:vext-indirect-water} reports the relative RMSE for the subset of observables that are optimized.
Despite the severe basis reduction, the error in the predicted eigenvalues is reduced by approximately a factor of two compared to training directly on the def2-TZVP Fock matrix. While the errors in dipole moments and Löwdin charges are slightly larger than those obtained by direct def2-TZVP training, they reflect the limited expressivity of the current architecture.

\subsection{Fock matrix predictions for organic molecules}
Finally, we consider the prediction of the Fock matrices for the QM7b dataset, which poses a substantially more challenging example due to its chemical diversity, varying stoichiometries, and heterogeneous local environments. For the previously mentioned QM7b dataset, we compute the Fock, density, and overlap matrices at the LDA/def2-SVP and LDA/def2-TZVP levels, and use the cc-PVQZ basis to represent the external potential.
As density matrix predictions are highly sensitive to deviations from their Grassmannian structure, causing the derived observables to degrade rapidly, especially as matrix size increases (larger basis, number of atoms in the atomic configuration), we restrict our examples to models where the target is the Fock matrix.

\subsubsection{Op2Op models}
Rather than directly predict the full Fock matrix $\mbf{H}$, we target only the nonlocal exchange-correlation correction, which is obtained as the difference of $\mbf{H}$ and the non-interacting Hamiltonian $\mbf{H}_\text{core} = \mbf{T} + \vext$, where $\mbf{T}$ is the kinetic energy operator. This decomposition allows the model to focus on learning the exchange-correlation effects, which are the most challenging to approximate, while the core Hamiltonian can be computed exactly and easily from the geometry through a non-self-consistent evaluation of one-electron integrals.

Even when the target is expressed in the minimal STO-3G basis, the blockwise decomposition requires almost 57 separate models of type~\eqref{eq:nonlinear-gating} to be trained for the possible values of $\boldsymbol{\gamma}$ corresponding to the combinations of species pairs and symmetry labels. 
A more expressive target basis, such as def2-SVP, increases this burden to 126, while the def2-TZVP basis representation requires 180 models to be trained, leading to a significant increase in the training complexity.  

For the STO-3G targets, the model achieves 0.37 eV error on the Fock matrix elements, 1.8 eV on the eigenvalues including virtual states, 0.34 eV on the HOMO-LUMO gaps, and 0.33 Debye on the dipole moment. The density matrix computed from the predicted Fock matrix incurs an RMSE of 3.5$\times 10^{-2}$ with respect to the reference values. 
When targeting the Fock matrix in the def2-SVP basis, on the other hand, a similar architecture achieves comparable accuracy on the Fock matrix elements with an RMSE of 0.55 eV, while the error on the density matrix computed from these predictions rises dramatically to the 1.5$\times 10^{2}$. This sharp increase underscores the strong sensitivity of the derived observables to small inaccuracies in the prediction of matrix elements owing to the nonlinear nature of the generalized eigenvalue problem and diagonalization~\eqref{eq:density-matrix}. 
\begin{figure}[hbp]
    \centering
    \includegraphics[width=1.0\linewidth]{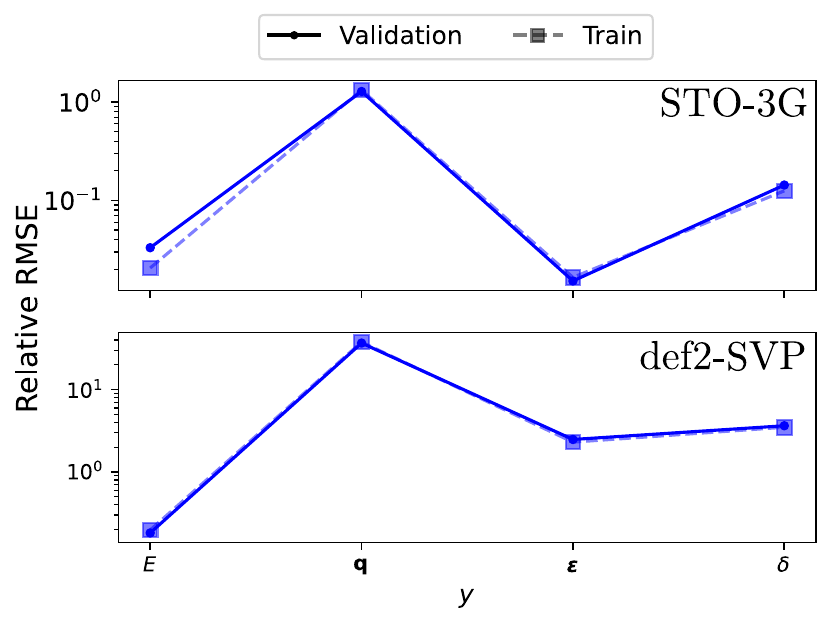}
    \caption{Relative RMSE on the total energy ($E$), L\"owdin charges ($\mbf{q}$), eigenvalues ($\boldsymbol{\varepsilon}$), and HOMO-LUMO gaps ($\delta$) normalized by the standard deviation of the reference values for the QM7 dataset. The observables are derived from the prediction of an \emph{Op2Op} model that was trained to optimize the elements of the Fock matrix in the STO-3G basis (top) and def2-SVP basis (bottom). In both cases, the model inputs were the symmetry-adapted blocks up to the third power of the $\vext$ expressed on the cc-PVQZ basis.}
    
    \label{fig:vext-direct-qm7}
\end{figure}

For a larger basis, the performance decreases further. In the most challenging case, when the target Fock matrix is expressed in the def2-TZVP basis, we observe the highest errors of 0.75 eV per matrix element of the Fock matrix. The HOMO-LUMO gap predictions similarly become worse with basis size, with RMSE values of around 4.4 eV error in both the SVP and TZVP representations. 

\begin{figure}
    \centering
    \includegraphics[width=1.0\linewidth]{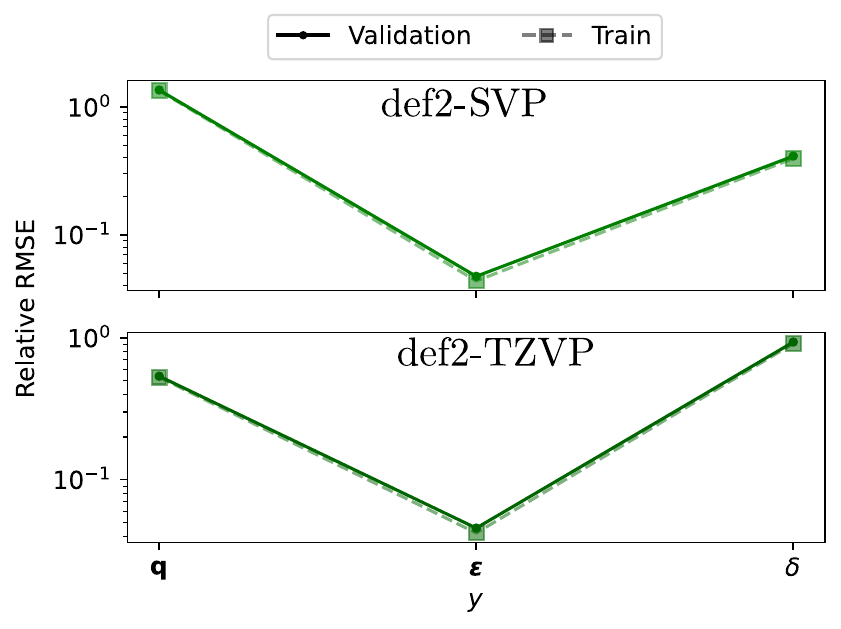}
    \caption{Relative RMSE on L\"owdin charges ($\mbf{q}$), eigenvalues ($\boldsymbol{\varepsilon}$) and HOMO-LUMO gaps ($\delta$) from the \emph{effective Op2Op} model. The reference observable values were computed using the $\mcal{B}_\mbf{y}=$ def2-SVP basis set (top) and def2-TZVP (bottom). The model predicts the downfolded Fock matrix in the $\mcal{B}_\mbf{M}=$ STO-3G basis in each case. The input $\vext$ in the cc-PVQZ basis, with matrix powers ~\eqref{eq:matrix-product} truncated at $\kappa_\text{max}=3$. The RMSEs have been normalized by the standard deviation of each observable.}
    \label{fig:vext-indirect-qm7}
\end{figure}

\subsubsection{Effective Op2Op models}
The benefit of the effective \emph{Op2Op} models becomes most apparent for this example. As previously observed, for identical $\vext$ inputs and network architectures, the accuracy of \emph{Op2Op} models degrades systematically as the target basis expands from STO-3G to def2-TZVP, and derived observables are especially sensitive to these errors. Instead of targeting the Fock matrix in the bases of reference calculations ($\mcal{B}_\mbf{y}$), we learn an effective matrix in the fixed basis size corresponding to $\mcal{B}_\mbf{M}$ = STO-3G by first obtaining the analytical projection of the matrix from the original, larger $\mcal{B}_\mbf{y}$ basis in the STO-3G basis following SM Section~\ref{sisec:basis-projection}. Using these projected matrices as an initial point for training, we optimize the model to reproduce derived observables corresponding to a subset of reference eigenvalues and L\"owdin charges.

As shown in Fig.~\ref{fig:vext-indirect-qm7}, this approach leads to an improvement of nearly two orders of magnitude in the accuracy of the eigenvalue, while the RMSE of the HOMO–LUMO gap is reduced to approximately 0.5 eV. These results show that operating on a compact latent basis and optimizing projected observables help mitigate the error amplification observed in the direct \emph{Op2Op} Fock predictions in larger bases. 

Finally, we note that these RMSEs are obtained without extensive architectural tuning and could be further improved.

\section{Discussion and Conclusions}
Although the external potential formally describes the Coulomb attraction between electrons and nuclei, when expressed on a basis of atom-centered orbitals, its representation $\vext$ is localized. This local structure leads to several similarities with atom-centered descriptors and provides a route to widely used atomistic descriptors from a shared underlying operator. Each matrix element $\vext^{\alpha\beta}(A_{ij})$ encodes one-neighbor correlations around the pair of atoms $(i,j)$, corresponding to a three-body interaction, much like body-ordered expansions used in traditional descriptors such as SOAP, ACE, and $N$-atom-centered density correlations. Likewise, the symmetry-adapted decomposition of $\vext$ into irreducible representations of $O(3)$ parallels the symmetrization procedures used for atom-centered density expansions. At the same time, we show that with a specific choice of basis, $\vext$ relates to the Coulomb matrix, indicating that descriptors often viewed as global can, in fact, be decomposed into local contributions depending on the representation used to express them. 

The matrix form of $\vext$ provides a particularly favorable and straightforward way to implement equivariant message passing through simple repeated matrix multiplications, bypassing the need for symmetrization at every message-passing step. Eq.~\eqref{eq:matrix-product} mimics the operations underlying message-passing graph neural networks, with messages passed between edges (matrix blocks) rather than atoms, but at a much reduced cost. In the product form, every atom interacts with all others. However, a notion of local neighborhoods can be enforced by zeroing matrix elements for atom pairs separated beyond a cutoff, providing a closer connection to equivariant GNNs that operate on local neighborhoods. This local-global duality of $\vext$ not only simplifies the construction and operations underlying popular descriptors and GNNs, but also allows it to overcome the limitations of strictly local descriptions and finite basis set discretizations that affect atom-centered descriptors, and describe nonlocal behavior. 

In this work, we propose a framework to predict both molecular properties and electronic structure operators from the external potential operator and refer to these approaches as \emph{Op2Prop} and \emph{Op2Op} models, respectively. 
The choice between these modeling approaches reflects a tradeoff between models optimized for specific property prediction and unified models that learn an electronic operator from which multiple properties can be derived. We demonstrate that linear \emph{Op2Prop} models based on $\vext$ match or exceed the performance of SOAP descriptors of comparable body order when predicting energies and dipole moments. 
In contrast, \emph{Op2Op} models aim to predict electronic operators that encapsulate a richer physical description of the system, but at the cost of increased complexity. Within this setting, we briefly explore the choice between targeting the Fock matrix $\mbf{H}$ and the density matrix $\mbf{P}$. Although these quantities are formally related, their matrix representations differ in conditioning and constraints, resulting in distinct learning problems in practice. Our results demonstrate that models trained on the Fock matrix consistently incur lower errors on molecular properties computed from them than models trained directly on the density matrix. We attribute the ease of learning the Fock matrix to its more local nature compared to the density matrix, as well as to the physical and mathematical constraints on $\mbf{P}$ that make it particularly sensitive to numerical errors. A deeper investigation of target choice, numerical stability, and downstream usability within electronic structure workflows will be an important direction for future work. 

We further demonstrate the applicability of \emph{Op2Op} models across multiple datasets, target basis sets, and supervision strategies. In terms of the latter, we compare models that target matrix elements of the operators in a specified basis set and models that infer an implicit effective representation through supervision of properties derived from it. We find that learning this compact latent basis representation provides a practical safeguard against nonlinear error amplifications during diagonalization of the generalized eigenvalue problem, especially when the target matrices are expressed in a large basis. 

All these examples point towards a broader integration of machine learning with electronic structure, which has traditionally been treated separately from descriptor design and modeling strategies. By incorporating several components of electronic structure as inputs, outputs, and intermediate quantities in the model workflow, our framework elevates basis sets and operators to tunable parameters of model design. 
Electronic structure theory provides a natural hierarchy of body-ordered quantities as three-center integrals correspond to three-body descriptors (as demonstrated and used in this work), two-center integrals (such as overlap or kinetic energy operators) to two-body contributions, and so on. Moreover, these quantities are routinely computed in any self-consistent field calculation and thus provide compact and physically meaningful model inputs at essentially no additional cost. Equally importantly, this allows atomistic machine learning models to leverage species-specific basis sets that have already been carefully optimized within quantum chemistry, instead of requiring the explicit construction of bespoke ML-specific bases~\cite{bigi+22jcp, darby2022compressing, goscinski2021optimal}, but the relative advantage of doing so remains an open question.

Although this work is motivated by the Hohenberg–Kohn theorems and centers on $\vext$ as the primary input, it naturally extends to alternative inputs and targets that encode the same underlying physics but may be numerically easier to learn, including operators other than Fock and density matrices,  analytic matrix functions, or spectral observables.

Finally, the connection between matrix products and message passing suggests that related ideas could be transferred back to geometry-based message-passing schemes, enabling efficient higher-order message passing without the cost of explicit symmetrization at every update step, which \rev{has been briefly explored~\cite{batatia2023equivariant}} and is a direction we plan to pursue in future work.

\section*{Supplementary Material}
Detailed derivations, additional experiments, as well as details and tabulated results for some figures in the main text, are reported in the accompanying supplementary material. 

\begin{acknowledgments}
JN is grateful for funding from the MIT Postdoctoral Fellowship for Excellence in Engineering and thanks the MIT Center for Computational Science and Engineering and the Schwarzman College of Computing for support. 
JN and TS acknowledge support from the National Science Foundation under Cooperative Agreement PHY-2019786 (The NSF AI Institute for Artificial Intelligence and Fundamental Interactions) and the Air Force Office of Scientific Research under Award No. FA9550-24-1-0067.
GD acknowledges support from the ANR project NUMERIQ (ANR-24-CE46-2255), as well as the support of the EIPHI Graduate school (contract ANR-17-EURE-0002) and the R\'egion 
Bourgogne-Franche-Comt\'e.
We thank Andrey Bryutkin and Youssef Marzouk for insightful discussions.

\noindent
\textbf{Conflict of interest statement}
The authors have no competing interests to declare.\\
\textbf{Data and materials availability} All the code used in this study will be made publicly available on publication.
\\
\textbf{Author contributions}
JN conceived the project and designed the research methodology and code infrastructure. GD and JN discussed the mathematical foundations of the methods as well as the overall research direction. JN wrote the first draft of the manuscript. All authors discussed the results and reviewed the final version of the manuscript.
\end{acknowledgments}

\bibliography{refs}
\onecolumngrid\clearpage


\section*{Supplementary material}

# Machine learning electronic structure and atomistic properties from the external potential
## Supplementary Material

Jigyasa Nigam,[1,*] Tess Smidt,[1] and Geneviève Dusson[2]

[1]*Research Laboratory of Electronics, Massachusetts Institute of Technology, Cambridge, 02139 MA, United States of America*

[2]*Université Marie et Louis Pasteur, CNRS, LmB (UMR 6623), F-25000 Besançon, France*

## I. NOTATION

In the text, we use $\mathbf{x}$ to denote general spatial coordinates, $\mathbf{r}$ to denote electronic coordinates, and $\mathbf{R}$ to denote nuclear coordinates. The matrix elements associated with orbital $\chi_\alpha$ centered on atom $i$, $\chi_\beta$ centered on atom $j$ in structure $A$, is denoted by a shorthand $\mathbf{M}^{\alpha\beta}(A_{ij})$. $\chi$ denotes atom-centered orbitals, while $\alpha$ and $\beta$ are triplets denoting the radial and angular quantum numbers $(nlm)$. The atomic number or chemical element corresponding to atom $i$ is denoted by $Z_i$. When referring to other atoms that do not explicitly index the matrix elements, we use the index $a$.

## II. REPRESENTATION OF THE EXTERNAL POTENTIAL

The nuclear potential in Eq. (1) can also be defined as,

$$\mathbf{V}(A) = \sum_{a \in A} \mathbf{V}_a(A). \tag{S1}$$

In an atom-centered basis set, the matrix elements of each summand contributing to (S1), $\mathbf{V}_a(A) = \frac{-Z_a}{|\mathbf{r}-\mathbf{R}_a|}$, corresponding to the atomic orbital $\chi_\alpha$ on atom $i$, $\chi_\beta$ on atom $j$, where $\alpha, \beta$ are shorthand for $(n_i l_i m_i)$ and $(n_j l_j m_j)$ respectively, are given by Eq. (2) reproduced below,

$$\mathbf{V}_a^{\alpha\beta}(A_{ij}) = -Z_a \int \frac{\chi_\alpha^*(\mathbf{r}-\mathbf{R}_i)\chi_\beta(\mathbf{r}-\mathbf{R}_j)}{|\mathbf{r}-\mathbf{R}_a|} d\mathbf{r}.$$

These matrix elements depend simultaneously on three nuclear positions $\mathbf{R}_a, \mathbf{R}_i, \mathbf{R}_j$. Each integral contributing to the sum depends on the triangular atomic geometry formed by $a, i, j$, and is equivalent to a three-body term. By performing the sum over $a$, the matrix element $\langle \chi_\alpha | \hat{V} | \chi_\beta \rangle$, is instead indexed by fixed bra center $i$, the ket center $j$, while the remaining nuclear dependence on $a$ enters through a linear superposition of the Coulomb fields generated by the surrounding nuclei, making it equivalent (but not equal) to a pair-descriptor $\rho_{ij}^{\otimes 1}$ proposed in Ref. 1.

Equivalently, we may write the matrix element

$$\mathbf{V}^{\alpha\beta}(A_{ij}) = \int d\mathbf{r}\, \chi_\alpha^*(\mathbf{r}-\mathbf{R}_i)\chi_\beta(\mathbf{r}-\mathbf{R}_j) \left[ \sum_{a \in A} \frac{-Z_a}{|\mathbf{r}-\mathbf{R}_a|} \right] \tag{S2}$$

as an object indexed by the pair of atoms $i, j$, whose dependence on the remaining "neighbor" nuclei is linear. Unlike the standard pair descriptor, this object does not explicitly depend on the relative vector $\mathbf{R}_i - \mathbf{R}_j$ but instead probes the environment of the pair through the Coulomb kernel.

---

* jnigam@mit.edu

Tensor products of the external-potential matrix elements generate higher-order dependence on the "neighbor" nuclei.

$$\mathbf{V}^{\alpha\beta}(A_{ij}) \otimes \mathbf{V}^{\alpha'\beta'}(A_{i'j'}) = \int d\mathbf{r}\, d\mathbf{r}' \sum_{a_1 a_2} \frac{\chi_\alpha^*(\mathbf{r}-\mathbf{R}_i)\chi_\beta(\mathbf{r}-\mathbf{R}_j)}{|\mathbf{r}-\mathbf{R}_{a_1}|} \frac{\chi_{\alpha'}^*(\mathbf{r}'-\mathbf{R}_{i'})\chi_{\beta'}(\mathbf{r}'-\mathbf{R}_{j'})}{|\mathbf{r}'-\mathbf{R}_{a_2}|}. \quad (S3)$$

Even restricting the tensor product to only orbital indices, such that $i'=i$ and $j'=j$, we obtain

$$\mathbf{V}^{\alpha\beta}(A_{ij}) \otimes \mathbf{V}^{\alpha'\beta'}(A_{ij}) = \int d\mathbf{r}\, d\mathbf{r}' \, \chi_\alpha^*(\mathbf{r}-\mathbf{R}_i)\chi_\beta(\mathbf{r}-\mathbf{R}_j)\chi_{\alpha'}^*(\mathbf{r}'-\mathbf{R}_i)\chi_{\beta'}(\mathbf{r}'-\mathbf{R}_j) \times$$
$$\sum_{a_1 a_2} \frac{1}{|\mathbf{r}-\mathbf{R}_{a_1}|} \frac{1}{|\mathbf{r}'-\mathbf{R}_{a_2}|} \quad (S4)$$

For a fixed pair $(i,j)$, this object depends quadratically on the surrounding nuclear density and therefore corresponds to a two-center, two-neighbor correlation. Higher tensor powers generate systematically higher-order pair-neighbor correlations, exactly as for many-body atom-centered density-based expansions.

On the other hand, matrix products can be written as,

$$\left(\mathbf{V}^2\right)^{\alpha\beta}(A_{ij}) = \sum_{a\eta} \mathbf{V}^{\alpha\eta}(A_{ia})\mathbf{V}^{\eta\beta}(A_{aj}), \quad (S5)$$

where $\eta$ indexes the orbital centered on the summed atom index $a$. This corresponds to a contraction of indices within the tensor product, as explained in the next section.

## A. Link to body-ordered descriptors

For comparison with geometric body-ordered descriptors, it is instructive to derive the atom-centered density correlations as follows. Writing $\mathbf{R}_{ai} = \mathbf{R}_a - \mathbf{R}_i$, the local atomic densities take the form,

$$\rho_i(\mathbf{x}) = \sum_{a \in \mathcal{N}_i} g(\mathbf{x}-\mathbf{R}_{ai}) = \int d\mathbf{R}\, \rho(\mathbf{R})\, g(\mathbf{x}-(\mathbf{R}-\mathbf{R}_i)), \quad (S6)$$

where we have defined the local neighborhood of atom $i$ as $\mathcal{N}_i$ and the *neighbor density field* $\rho(\mathbf{R})$

$$\rho(\mathbf{R}) = \sum_{a \in \mathcal{N}_i} \delta(\mathbf{R}-\mathbf{R}_a). \quad (S7)$$

Projecting the atom density in Eq. (S6) on a basis of radial and angular functions, we obtain density coefficients $c_{nlm}(A_i)$, for the neighbor density centered on atom $i$ in the configuration $A$ (denoted

as $A_i$),

$$c_{nlm}(A_i) = \int d\mathbf{x}\, R_{nl}(x) Y_l^m(\hat{\mathbf{x}}) \int d\mathbf{R}\, \rho(\mathbf{R})\, g(\mathbf{x} - (\mathbf{R} - \mathbf{R}_i)) \tag{S8}$$

$$= \int d\mathbf{R}\, \rho(\mathbf{R})\, f_{nlm}^i(\mathbf{R}) \tag{S9}$$

$$= \sum_{a \in \mathcal{N}_i} f_{nlm}^i(\mathbf{R}_a) \tag{S10}$$

where we interchanged the order of integration in the second line, $f_{nlm}^i(\mathbf{R}) = \int d\mathbf{x}\, R_{nl}(x) Y_l^m(\hat{x}) g(\mathbf{x} - (\mathbf{R} - \mathbf{R}_i))$.

As Eq. (S9) is linear neighbor density, it describes the correlations of the central atom $i$ with a single neighbor. To obtain second-order correlations, equivalent to the power spectrum, we consider the tensor product $c_{nlm}(A_i) \otimes c_{n'l'm'}(A_i)$ Using Eq. (S9),

$$\rho_i^{\otimes 2} = \int d\mathbf{R} \int d\mathbf{R}'\, f_{nlm}^i(\mathbf{R}) \rho(\mathbf{R}) f_{n'lm}^i(\mathbf{R}') \rho(\mathbf{R}'). \tag{S11}$$

Symmetrizing the tensor product to invariants, which from the rules of angular momentum coupling yields $l = l'$ and $m = m'$ (real-valued spherical harmonics), we recover the power spectrum a two-point correlation of the atom density centered on $i$

$$\rho_i^{\otimes 2, \lambda = 0} = \sum_m \sum_{a_1 a_2 \in \mathcal{N}_i} f_{nlm}^i(\mathbf{R}_{a_1}) f_{n'lm}^i(\mathbf{R}_{a_2}),$$

The superscript $\lambda = 0$ on the LHS denotes symmetrization with respect to rotations, and the quantity on the right hand side corresponds to the set of all unordered triplet interactions involving the central atom $i$. When it is clear from context that we are referring to rotational invariants, we will drop the superscript and denote the symmetrized density as $\rho_i^{\otimes 2}$ (as in the main text).

Higher-order correlations of the atom-density may be similarly obtained by symmetrizing the tensor product of $\nu$ copies of the atom-density [2], and correspond to $\nu$-point correlation functions of the atom-centered density [3]. The global descriptor for a structure $A$ can be obtained by summing over the contributions S11 from all atoms in the structure,

$$\rho^{\otimes 2} = \sum_{i \in A} \rho_i^{\otimes 2}. \tag{S12}$$

We can now analogously interpret the elements of the external potential matrices and their products. If we generalize the neighbor density field $\rho(\mathbf{R})$ in Eq. (S7) to reflect atomic charges, as the *charge*

4

*density field*

$$\rho_Z(\mathbf{R}) = \sum_a Z_a \delta(\mathbf{R} - \mathbf{R}_a),$$

the external potential matrix element in Eq. (S2) can be rewritten as

$$\mathbf{V}^{\alpha\beta}(A_{ij}) = \int d\mathbf{R}\, \rho_Z(\mathbf{R}) F^{ij}_{\alpha\beta}(\mathbf{R}), \tag{S13}$$

where $F^{ij}_{\alpha\beta}(\mathbf{R}) = -\int d\mathbf{r} \frac{\chi^*_\alpha(\mathbf{r}-\mathbf{R}_i)\chi_\beta(\mathbf{r}-\mathbf{R}_j)}{|\mathbf{r}-\mathbf{R}|}$. For fixed orbital indices $\alpha, \beta$ and pair centers $i, j$, the matrix element is linear in charge (i.e. weighted atomic) density and represents the one-neighbor correlation between the atom-pair and its environment. Unlike the atomic descriptors that correlate the atom-density with local functions $f^i_{nlm}$, the external potential correlates the charge density with a nonlocal Coulomb kernel weighted by the overlap of the basis functions centered on the two atoms. Tensor products of $\mathbf{V}^{\alpha\beta}$ correspond to higher-order correlations of the charge density, generalizing Eq. (S4) to

$$\mathbf{V}^{\otimes\nu} \sim \int d\mathbf{R}_1 \ldots d\mathbf{R}_\nu\, \rho_Z(\mathbf{R}_1) \ldots \rho_Z(\mathbf{R}_\nu) \prod_{k=1}^{\nu} F^{ij}_{\alpha_k \beta_k}(\mathbf{R}_k), \tag{S14}$$

which are systematically increasing body-ordered expansions of the pair, analogous to the $\nu$-point correlation functions of the atom-centered densities.

On the other hand, the matrix product of $\mathbf{V}$ (S5) can be mapped to a specific form of the product above, coupled through a shared intermediate atomic center

$$\left(\mathbf{V}^2\right)^{\alpha\beta}(A_{ij}) = \sum_{a\eta} \int d\mathbf{R} d\mathbf{R}'\, \rho_Z(\mathbf{R})\rho_Z(\mathbf{R}') F^{ij}_{\alpha\eta}(\mathbf{R}) F^{aj}_{\eta\beta}(\mathbf{R}').$$

### B. Link to Coulomb matrix

Consider the definition of the matrix element in Eq. (S2), and the Dirac-delta limit of local basis sets resulting in a basis set with exactly one function per chemical species, such that $\chi_\alpha(\mathbf{r}-\mathbf{R}) = \chi_\beta(\mathbf{r}-\mathbf{R}) = \delta(\mathbf{r}-\mathbf{R})$. Then the matrix elements evaluate to,

$$\mathbf{V}(A_{ij}) = \int d\mathbf{r}\, \delta(\mathbf{r}-\mathbf{R}_i)\delta(\mathbf{r}-\mathbf{R}_j) \left[\sum_a \frac{-Z_a}{|\mathbf{r}-\mathbf{R}_a|}\right] \tag{S15}$$

Using $\int d\mathbf{r}\, \delta(\mathbf{r}-\mathbf{R}_i)\delta(\mathbf{r}-\mathbf{R}_j)f(\mathbf{r}) = \delta(\mathbf{R}_i-\mathbf{R}_j)f(\mathbf{R}_i)$, we obtain

$$\mathbf{V}(A_{ij}) = \sum_a \frac{-Z_a}{|\mathbf{R}_i - \mathbf{R}_a|} \delta_{ij}. \tag{S16}$$

Compared to the definition of the matrix elements of the Coulomb matrix CM between atoms $i$ and $j$,

$$\text{CM}(A_{ij}) = \frac{Z_i Z_j}{|\mathbf{R}_i - \mathbf{R}_j|}, \tag{S17}$$

the dirac-Delta basis representation of the external potential thus corresponds to a row or column sum of the Coulomb matrix, up to a factor of $Z_i$, $\sum_a \text{CM}(A_{ai}) = \sum_a \frac{Z_i Z_a}{|\mathbf{R}_a - \mathbf{R}_i|}$.

A local AO representation of $\mathbf{V}$ generalizes the Coulomb matrix from point charges to a localized charge distribution, which can be then expanded as multipoles.

## C. Link to message passing

As will be shown in Section III, matrix products of the form Eq. (S5), expanded as Eq. (S23), propagate information between atom pairs through intermediate atomic centers. This structure naturally connects to message-passing neural networks (MPNNs) [4, 5], with the important distinction that in our framework, pairs of atoms (edges) rather than individual atoms (nodes) serve as the fundamental units.

Specifically, consider the matrix product $(\mathbf{M}^2)^{n_i l_i n_j l_j}_{m_i m_j}(A_{ij})$ defined in Eq. (S23), which aggregates information from intermediate atom $a$:

$$(\mathbf{M}^2)^{n_i l_i n_j l_j}_{m_i m_j}(A_{ij}) = \sum_{a n_a l_a m_a} \mathbf{M}^{n_i l_i n_a l_a}_{m_i m_a}(A_{ia}) \mathbf{M}^{n_a l_a n_j l_j}_{m_a m_j}(A_{aj}). \tag{S18}$$

This can be interpreted as a message-passing operation where

- The matrix elements $\mathbf{M}^{n_i l_i n_a l_a}_{m_i m_a}(A_{ia})$ and $\mathbf{M}^{n_a l_a n_j l_j}_{m_a m_j}(A_{aj})$ represent edge features (messages) associated with pairs $(i, a)$ and $(a, j)$ respectively.

- The sum over intermediate atom $a$ and its orbital indices $(n_a l_a m_a)$ aggregates messages from all neighbors of the edge $(i, j)$. If no distance cutoff is imposed on the matrix elements (meaning elements beyond a given separation are not set to zero), this aggregation effectively spans the entire structure.

- The orbital labels $(n_i l_i m_i, n_j l_j m_j)$ play the role of equivariant feature indices that can be manipulated internally but do not define the graph topology, analogous to node/edge features in equivariant MPNNs.

6

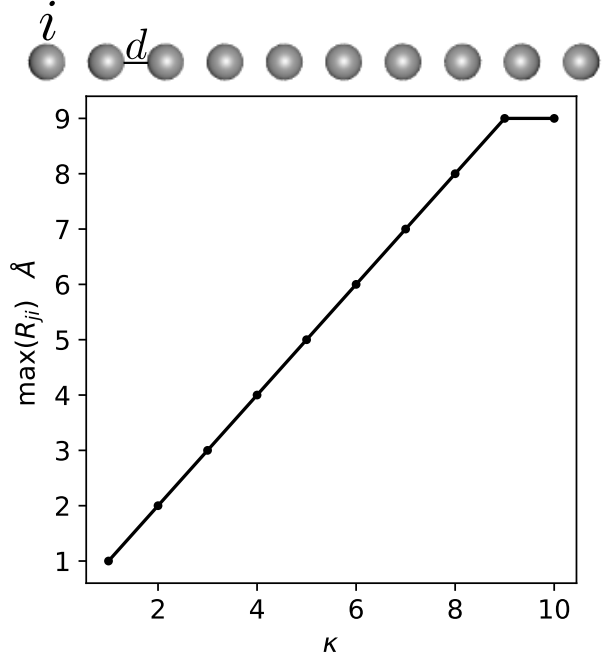

FIG. S1: For a linear chain of 10 carbon atoms with an interatomic separation of d = 1 Å, we compute **V** using the def2-TZVP basis, and impose an explicit cutoff of 1 Å, i.e. set all elements corresponding to atoms beyond nearest neighbor to zero. At matrix power $\kappa = 1$, the receptive field (the distance up to which a particular atom can see another) is therefore 1 Å. However, successive matrix products populate matrix elements such that the receptive field increases linearly upto the maximum distance.

Successive matrix powers correspond to multi-step message passing, with each multiplication propagating information through an additional intermediate atomic center. As shown in Section III, the orthogonality relation in Eq. (S26) ensures that matrix products maintain rotational equivariance while aggregating information through rotational invariants at intermediate centers. The effective receptive field increases linearly with the number of message-passing steps (each matrix multiplication corresponds to an effective increase by one-hop neighborhood [6, 7]), as illustrated in Fig. S1.

## III. EQUIVARIANCE OF MATRIX PRODUCTS

Denote the matrix associated with structure $A$ as $\mathbf{M}(A)$ and the corresponding matrix obtained through rotations of the underlying structure as $\mathbf{M}(\hat{R}A)$. The matrix transforms under rotation as $\mathbf{M}(\hat{R}A) = \hat{R}\mathbf{M}(A)\hat{R}^\dagger$, which when expanded in terms of the matrix elements $\mathbf{M}_{m_i,m_j}^{n_i l_i n_j l_j}(A_{ij}) \equiv$

$\langle i\, n_i l_i m_i |\, \hat{M}\, | j\, n_j l_j m_j \rangle$ can be written as,

$$\mathbf{M}_{m_i m_j}^{n_i l_i n_j l_j}((\hat{R}A)_{ij}) = \sum_{m'_i m'_j} D^{(l_i)}_{m_i m'_i}(\hat{R}) \mathbf{M}_{m'_i m'_j}^{n_i l_i n_j l_j}(A_{ij}) D^{\dagger(l_j)}_{m'_j m_j}(\hat{R}). \tag{S19}$$

The Kronecker product of Wigner-D matrices can be reduced to a direct sum as

$$D^{(l_i)}_{m_i m'_i}(\hat{R}) D^{\dagger(l_j)}_{m'_j m_j}(\hat{R}) = \sum_{\lambda=|l_i-l_j|}^{(l_i+l_j)} \sum_{\mu\mu'=-\lambda}^{\lambda} D^{\lambda}_{\mu\mu'}(\hat{R}) \langle l_i m_i; l_j m_j | \lambda \mu \rangle \langle l_i m'_i; l_j m'_j | \lambda \mu' \rangle \delta_{\mu, m_i+m_j} \delta_{\mu', m'_i+m'_j},$$

separating the rotation of the matrix into rotations of irreps. The Clebsch-Gordan (CG) coefficients $\langle l_i m_i; l_j m_j | \lambda \mu \rangle$ uniquely fix the coupled magnetic quantum number, denoted by the Dirac-deltas on the RHS above. Each irrep is obtained by transforming the matrix elements from $\mathbf{M}_{m_i m_j}^{n_i l_i n_j l_j}(A_{ij})$ (uncoupled / unsymmetrized basis representation) to $\mathbf{M}_{\mu}^{\mathbf{b},\lambda}(A_{ij})$ (coupled / symmetrized basis representation) using Clebsch–Gordan coefficients for the transformation,

$$\mathbf{M}_{m_i m_j}^{n_i l_i n_j l_j}(\hat{R} A_{ij}) = \sum_{\lambda=|l_i-l_j|}^{(l_i+l_j)} D^{\lambda}_{\mu\mu'}(\hat{R}) \langle l_i m_i; l_j m_j | \lambda \mu \rangle \tag{S20}$$

$$\times \left( \sum_{m'_i m'_j} \langle l_i m'_i; l_j m'_j | \lambda \mu' \rangle \mathbf{M}_{m'_i m'_j}^{n_i l_i n_j l_j}(A_{ij}) \right) \tag{S21}$$

$$= \sum_{\lambda=|l_i-l_j|}^{(l_i+l_j)} D^{\lambda}_{\mu\mu'}(\hat{R}) \mathbf{M}_{\mu'}^{n_i l_i n_j l_j, \lambda}(A_{ij}) \langle l_i m_i; l_j m_j | \lambda \mu \rangle \tag{S22}$$

where in the last line $\langle l_i m_i; l_j m_j | \lambda \mu \rangle$ transforms the coupled representation back to the uncoupled form.

Now, consider the definition of the product $\mathbf{M}^2$

$$(\mathbf{M}^2)_{m_i m_j}^{n_i l_i n_j l_j}(A_{ij}) = \sum_{a n_a l_a m_a} \mathbf{M}_{m_i m_a}^{n_i l_i n_a l_a}(A_{ia}) \mathbf{M}_{m_a m_j}^{n_a l_a n_j l_j}(A_{aj}). \tag{S23}$$

Under rotations, using (S19), this product transforms as,

$$(\mathbf{M}^2)_{m_i m_j}^{n_i l_i n_j l_j}(\hat{R} A_{ij}) = \sum_{a n_a l_a m_a} \sum_{m'_i m'_a m''_a m'_j} D^{(l_i)}_{m_i m'_i}(\hat{R}) \mathbf{M}_{m'_i m'_a}^{n_i l_i n_a l_a}(A_{ia}) D^{\dagger(l_a)}_{m'_a m_a}(\hat{R}) \tag{S24}$$

$$\times D^{(l_a)}_{m_a m''_a}(\hat{R}) \mathbf{M}_{m''_a m'_j}^{n_a l_a n_j l_j}(A_{aj}) D^{\dagger(l_j)}_{m'_j m_j}(\hat{R}) \tag{S25}$$

Using orthogonality of Wigner matrices,

$$\sum_{m_a} D^{\dagger(l_a)}_{m'_a m_a}(\hat{R}) D^{(l_a)}_{m_a m''_a}(\hat{R}) = \delta_{m'_a m''_a}, \tag{S26}$$

8

we obtain

$$(\mathbf{M}^2)_{m_i m_j}^{n_i l_i n_j l_j}(\hat{R} A_{ij}) = \sum_{m'_i m'_j} D_{m_i m'_i}^{(l_i)}(R) D_{m'_j m_j}^{\dagger(l_j)}(R) \tag{S27}$$

$$\left( \sum_{a n_a l_a m'_a} \mathbf{M}_{m'_i m'_a}^{n_i l_i n_a l_a}(A_{ia}) \mathbf{M}_{m'_a m'_j}^{n_a l_a n_j l_j}(A_{aj}) \right) \tag{S28}$$

$$= \sum_{m'_i m'_j} D_{m_i m'_i}^{(l_i)}(R) (\mathbf{M}^2)_{m'_i m'_j}^{n_i l_i n_j l_j}(A_{ij}) D_{m'_j m_j}^{\dagger(l_j)}(R), \tag{S29}$$

where we used the definition of the matrix product (S23). Following the steps as in Eq. (S20), $\mathbf{M}^2$ transforms the same way as $\mathbf{M}(A)$ under rotations, and this formulation generalizes to higher matrix products.

The orthogonality relation (S26) arises from the unitarity of rotations, and is effectively equivalent to projecting the tensor product of the orbitals associated with the summed intermediate atomic center $a$ to invariants. We can see this by starting from the definition of the tensor product (15) and suppressing radial quantum numbers for clarity,

$$(\mathbf{M}^{\otimes 2})_M^{\mathbf{\Gamma}} = \sum_{m_i m_{a_1} m_{a_2} m_j} \mathbf{M}_{m_i m_{a_1}}^{l_i l_{a_1}}(A_{i a_1}) \mathbf{M}_{m_{a_2} m_j}^{l_{a_2} l_j}(A_{a_2 j}) \times$$

$$\langle l_i m_i; l_{a_1} m_{a_1} | \lambda_1 \mu_1 \rangle \, \langle l_{a_2} m_{a_2}; l_j m_j | \lambda_2 \mu_2 \rangle \, \langle \lambda_1 \mu_1; \lambda_2 \mu_2 | \Lambda M \rangle, \tag{S30}$$

The structure of the Clebsch–Gordan coefficients in the expression above corresponds to the angular momentum coupling scheme

$$(l_i \otimes l_{a_1}) \to \lambda_1, \qquad (l_{a_2} \otimes l_j) \to \lambda_2, \qquad (\lambda_1 \otimes \lambda_2) \to \Lambda.$$

That is, the orbital angular momenta associated with the pairs $A_{i a_1}$ and $A_{a_2 j}$ are first coupled independently to intermediate irreducible representations $\lambda_1$ and $\lambda_2$, which are then combined to form the total angular momentum $\Lambda$.

Using angular momentum recoupling [8], the same tensor product can be equivalently expressed in a different coupling order,

$$(l_{a_1} \otimes l_{a_2}) \to \lambda'_1, \qquad (l_i \otimes l_j) \to \lambda'_2, \qquad (\lambda'_1 \otimes \lambda'_2) \to \Lambda,$$

where the two angular momenta associated with the intermediate atomic center are coupled first. This change of coupling order is implemented by standard Wigner-$6j$ recoupling and leaves the overall content of Eq. (S30) unchanged.

Restricting this tensor product to $l_{a_1} = l_{a_2} = l_a$ and denoting $m_{a_1}$ as $m_a$, $m_{a_2}$ and $m'_a$ and changing the order of symmetrization

$$(\mathbf{M}^{\otimes 2})^{\mathbf{\Gamma}}_M \propto \sum_{m_i m_a m'_a m_j} \langle l_a m_a; l_a m'_a | \lambda'_1 \mu'_1 \rangle \, \langle l_i m_i; l_j m_j | \lambda'_2 \mu'_2 \rangle \, \langle \lambda'_1 \mu'_1; \lambda'_2 \mu'_2 | \Lambda M \rangle \, \mathbf{M}^{(l_i l_a)}_{m_i m_a}(A_{ia}) \mathbf{M}^{(l_a l_j)}_{m'_a m_j}(A_{aj}), \quad (S31)$$

where the constant of proportionality is fixed by the Wigner-6j symbol.

If we now consider only $\lambda'_1 = 0$, and (for real spherical harmonics) use

$$\langle l_a m_a; l_a m'_a | 0\, 0 \rangle \propto \delta_{m_a m'_a},$$

we obtain

$$(\mathbf{M}^{\otimes 2})^{\Gamma} \propto \sum_{m_i m_j} \left( \sum_{m_a} \mathbf{M}^{(l_i l_a)}_{m_i m_a}(A_{ia}) \mathbf{M}^{(l_a l_j)}_{m_a m_j}(A_{aj}) \right) \langle l_i m_i; l_j m_j | \Lambda M \rangle. \quad (S32)$$

The term enclosed in the parenthesis is precisely the matrix product in Eq. (S23), followed by the usual symmetrization on the matrix elements into irreps as in Eq. (5). Thus, the matrix product corresponds to propagating invariants through the intermediate centers.

## IV. MATRIX ELEMENTS AS A FUNCTION OF DISTANCE

The evaluation of the nuclear attraction matrix elements involves factorizing the integral into an exponentially decaying Gaussian overlap and a Boys [9–11] function. Furthermore, all practical implementations employ an adaptive numerical cutoff over which the integral is performed.

For the QM7 dataset, truncating the matrix elements at 6 Å leads to a difference (RMSE) in $\mathbf{H}_{\text{STO-3G}}$ of 7.35 $\times 10^{-6}$ a.u. on the training set and 6.27 $\times 10^{-6}$ a.u. Therefore, this limits the accuracy attainable by any geometric model with a cutoff of 6 Å even at optimal fitting.

The impact of the same truncation is substantially more pronounced for the density matrix. Setting elements beyond $r_{ij} > 6$ Å incurs a deviation from the reference calculations $\mathbf{P}_{\text{STO-3G}}$ of 0.029 and 0.010 on the training and validation set, respectively. This reflects the intrinsically more delocalized nature of $\mathbf{P}$, also evident from Fig. S2. For larger and more diffuse basis sets, such as def2-TZVP, truncation effects are amplified more for the Fock matrix. At the same cutoff, the resulting errors with respect to the reference increase to approximately 7.11 $\times 10^{-3}$ a.u. and 1.24 $\times 10^{-3}$ a.u. on the train and validation splits of $\mathbf{H}_{\text{def2-TZVP}}$, and 0.046 and 0.010 on the train and validation splits

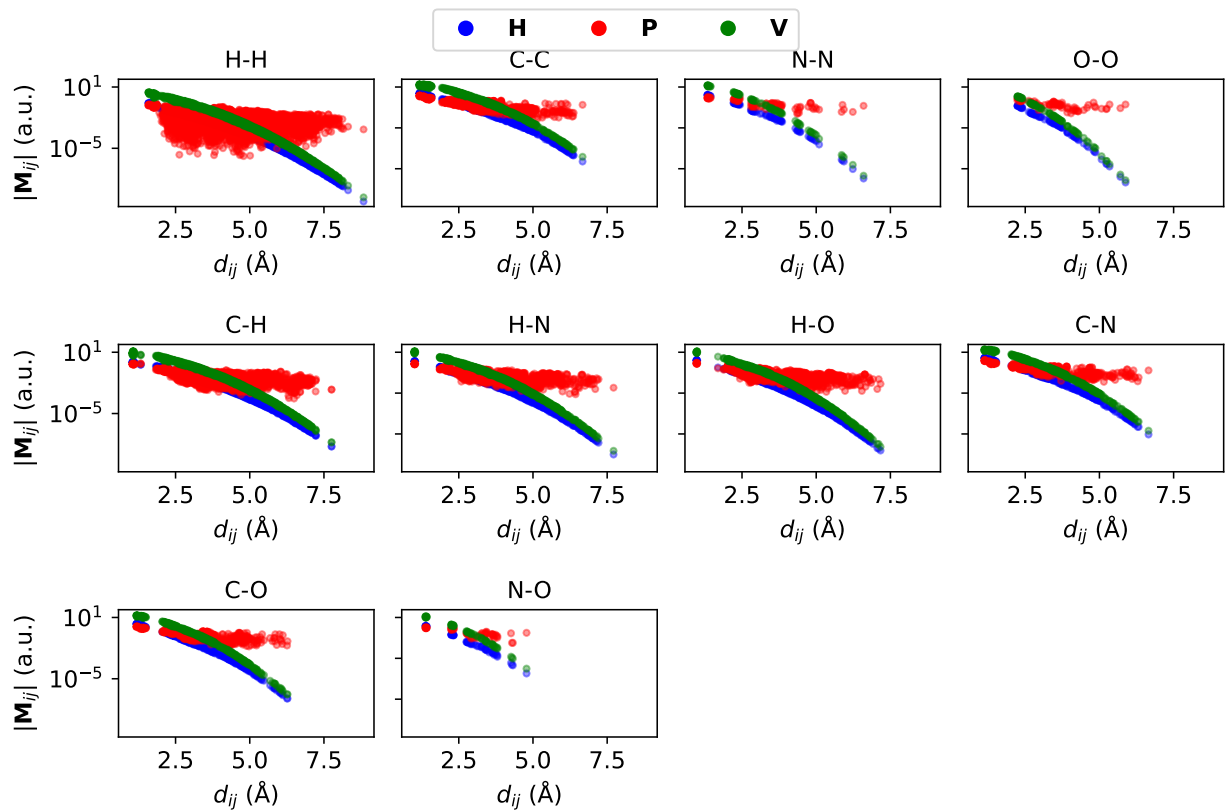

FIG. S2: Norm of the matrix elements $|\mathbf{M}_{ij}|$ in the QM7/STO-3G dataset as a function of the interatomic distance $d_{ij}$ between atoms $i$ and $j$ for various species pairs and for $\mathbf{M}$ corresponding to the Fock (blue), density matrix (red) and the external potential matrix (green).

of $\mathbf{P}_{\text{def2-TZVP}}$. Truncation affects $\mathbf{H}$ and $\mathbf{P}$ in qualitatively different ways. The matrix elements of $\mathbf{H}$ decay with interatomic distance as a consequence of the spatial decay of the basis function. In a more diffuse basis, the slower decay of orbitals implies non-negligible interactions even at larger interatomic separations $r_{ij}$, leading to an increasing truncation error for $\mathbf{H}$ with basis set. On the other hand, the nonlocality of the density matrix arises from the nonlocal algebraic constraints it satisfies (it is a projector onto the occupied subspace). The truncation errors in $\mathbf{P}$ correlate strongly with violations of the Grassmannian constraint ($PSP = 2P$) and are comparatively insensitive to the diffuseness of the basis.

Note that truncating $\mathbf{H}$ and recomputing the density matrix from the truncated Hamiltonian does not yield the same result as truncating $\mathbf{P}$ directly. The resulting projector no longer commutes with the original Hamiltonian, $[\mathbf{H}_{\text{truncated}}, \mathbf{P}_{\text{truncated}}] \neq 0$, which directly implies a violation of the Grassmannianity.

11

## V. ALGEBRAIC CONSTRAINTS ON THE MATRIX REPRESENTATIONS

For the electronic Hamiltonian $\mathbf{H}$ expressed on a non-orthogonal atomic orbital (AO) basis with overlap matrix $\mathbf{S}$, the generalized eigenvalue equation is given by Eq. (6) in the main text,

$$\mathbf{HC} = \mathbf{SC}\,\mathrm{diag}\,\boldsymbol{\varepsilon}, \tag{S33}$$

where each column of $\mathbf{C}$ contains the AO basis coefficients for a molecular orbital (MO), $\boldsymbol{\varepsilon}$ is the vector of corresponding eigenvalues (MO energies), and $\mathbf{S}$ is the overlap matrix of the AOs.

The density matrix is constructed from the eigenvectors and occupation numbers as given in Eq. (7) of the main text,

$$\mathbf{P} = \mathbf{C}\,\mathrm{diag}(\mathbf{f})\mathbf{C}^\dagger, \tag{S34}$$

where $\mathbf{f} \in \{0,1\}$ denotes the occupation of each MO. For a closed-shell single determinant, each occupied orbital has $f=1$ and unoccupied orbitals have $f=0$, so that $f^2 = f$.

We now prove that the eigenvectors satisfy the orthonormality condition $\mathbf{C}^\dagger \mathbf{SC} = \mathbf{1}$. Take two eigenvectors $\mathbf{c}_i$ and $\mathbf{c}_j$ (columns of $\mathbf{C}$) with eigenvalues $\varepsilon_i$ and $\varepsilon_j$. From the generalized eigenvalue equation (6),

$$\mathbf{Hc}_i = \varepsilon_i \mathbf{Sc}_i, \quad \mathbf{Hc}_j = \varepsilon_j \mathbf{Sc}_j.$$

Left-multiplying the first equation by $\mathbf{c}_j^\dagger$ gives

$$\mathbf{c}_j^\dagger \mathbf{Hc}_i = \varepsilon_i \mathbf{c}_j^\dagger \mathbf{Sc}_i.$$

Taking the Hermitian conjugate of the second equation and right-multiplying by $\mathbf{c}_i$, and using the fact that both $\mathbf{H}$ and $\mathbf{S}$ are Hermitian, we get

$$\mathbf{c}_j^\dagger \mathbf{Hc}_i = \varepsilon_j \mathbf{c}_j^\dagger \mathbf{Sc}_i,$$

Subtracting the two expressions gives

$$(\varepsilon_i - \varepsilon_j)\,\mathbf{c}_j^\dagger \mathbf{Sc}_i = 0.$$

For non-degenerate eigenvalues $\varepsilon_i \neq \varepsilon_j$, this implies

$$\mathbf{c}_j^\dagger \mathbf{Sc}_i = 0,$$

i.e., distinct eigenvectors are orthogonal. In the case of degeneracies, one may always choose linear combinations within the degenerate subspace such that this condition holds. Finally, each eigenvector can be normalized so that

$$\mathbf{c}_i^\dagger \mathbf{S} \mathbf{c}_i = 1.$$

Combining these conditions, we obtain $\mathbf{C}^\dagger \mathbf{S} \mathbf{C} = \mathbf{1}$.

Using this orthonormality condition, we can now prove that the density matrix satisfies the idempotency relation $\mathbf{PSP} = 2\mathbf{P}$. Substituting the definition of $\mathbf{P}$:

$$\mathbf{PSP} = (\mathbf{C}\operatorname{diag}(\mathbf{f})\mathbf{C}^\dagger)\mathbf{S}(\mathbf{C}\operatorname{diag}(\mathbf{f})\mathbf{C}^\dagger) \tag{S35}$$

$$= \mathbf{C}\operatorname{diag}(\mathbf{f})(\mathbf{C}^\dagger \mathbf{S} \mathbf{C})\operatorname{diag}(\mathbf{f})\mathbf{C}^\dagger \tag{S36}$$

$$= \mathbf{C}\operatorname{diag}(\mathbf{f})\mathbf{1}\operatorname{diag}(\mathbf{f})\mathbf{C}^\dagger \tag{S37}$$

$$= \mathbf{C}\operatorname{diag}(\mathbf{f}^2)\mathbf{C}^\dagger. \tag{S38}$$

Since $f^2 = f$ for $f \in \{0, 1\}$, we have $\operatorname{diag}(\mathbf{f}^2) = \operatorname{diag}(\mathbf{f})$, which would give $\mathbf{PSP} = \mathbf{P}$. However, in closed-shell systems, each occupied spatial orbital contains two electrons (one with spin up and one with spin down). The factor of 2 in the relation $\mathbf{PSP} = 2\mathbf{P}$ accounts for this double occupancy. For closed-shell systems, the occupation numbers in the density matrix construction should be interpreted as $f \in \{0, 2\}$ for unoccupied and occupied orbitals respectively, even though we use the notation $f \in \{0, 1\}$ to indicate occupation. Noting that for $f = 2$ we have $f^2 = 2f$, we obtain

$$\mathbf{PSP} = 2\mathbf{C}\operatorname{diag}(\mathbf{f})\mathbf{C}^\dagger = 2\mathbf{P}. \tag{S39}$$

For an orthogonal basis representation where $\mathbf{S} = \mathbf{1}$, the idempotency relation simplifies to $\mathbf{P}^2 = 2\mathbf{P}$. This can be verified by transforming to an Löwdin orthogonalized basis. Defining the orthogonalized density matrix $\overline{\mathbf{P}} = \mathbf{S}^{1/2}\mathbf{P}\mathbf{S}^{1/2}$, we have

$$\mathbf{S}^{1/2}\mathbf{PSPS}^{1/2} = 2\mathbf{S}^{1/2}\mathbf{P}\mathbf{S}^{1/2} \tag{S40}$$

$$(\mathbf{S}^{1/2}\mathbf{P}\mathbf{S}^{1/2})(\mathbf{S}^{1/2}\mathbf{P}\mathbf{S}^{1/2}) = 2\mathbf{S}^{1/2}\mathbf{P}\mathbf{S}^{1/2} \tag{S41}$$

$$\implies \overline{\mathbf{P}}^2 = 2\overline{\mathbf{P}}, \tag{S42}$$

where we have used $\mathbf{S}^{1/2}\mathbf{S}\mathbf{S}^{1/2} = \mathbf{S}^2 = \mathbf{1}$ in the orthogonal basis.

## VI. ADAPTIVE PROJECTION OF MATRICES ACROSS BASES

Let $\mathcal{B}_1$ and $\mathcal{B}_2$ be two non-orthogonal basis sets on which $\mathbf{M}$ is expressed. The two bases span Hilbert spaces $\mathcal{H}_1 = \text{span}(\mathcal{B}_1)$, and $\mathcal{H}_2 = \text{span}(\mathcal{B}_2)$. In the following, let $N_i$ denote the dimension, $\dim \mathcal{H}_i$, the AO's in $\mathcal{B}_1$ be $\left\{\chi_\mu^{(\mathcal{B}_1)}\right\}_{\mu=1}^{N_1}$ and $\mathcal{B}_2 = \left\{\chi_\mu^{(\mathcal{B}_2)}\right\}_{\mu=1}^{N_2}$, with $N_2 < N_1$ (i.e. we choose the bases such that $\dim \mathcal{H}_2 < \dim \mathcal{H}_1$).

The overlap matrices in the corresponding bases have elements specified by,

$$\mathbf{S}_{\mu\nu}^{(\mathcal{B}_1)} = \left\langle \chi_\mu^{(\mathcal{B}_1)} \mid \chi_\nu^{(\mathcal{B}_1)} \right\rangle, \quad \mathbf{S}_{\alpha\beta}^{(\mathcal{B}_2)} = \left\langle \chi_\alpha^{(\mathcal{B}_2)} \mid \chi_\beta^{(\mathcal{B}_2)} \right\rangle.$$

As the bases are non-orthogonal, the completeness relation for each basis $\mathcal{B}_i$ is given by

$$\sum_{\alpha\beta} |\chi_\alpha\rangle \left(\mathbf{S}^{(\mathcal{B}_i)}\right)^{-1}_{\alpha\beta} \langle\chi_\beta| = \mathbf{1} \tag{S43}$$

We further consider the cross-overlap of AOs in the two and define

$$S_{\alpha\mu}^{(\mathcal{B}_2\mathcal{B}_1)} = \left\langle \chi_\alpha^{(\mathcal{B}_2)} \mid \chi_\mu^{(\mathcal{B}_1)} \right\rangle. \tag{S44}$$

Let $\mathbf{M}$ be a self-adjoint operator, such as the Fock matrix. The generalized eigenproblem in basis $\mathcal{B}_i$ defined through the representation of the operator in this basis, $\mathbf{M}^{(\mathcal{B}_i)}$ and the overlap $\mathbf{S}^{(\mathcal{B}_i)}$, is

$$\mathbf{M}^{(\mathcal{B}_i)} \mathbf{C}^{(\mathcal{B}_i)} = \mathbf{S}^{(\mathcal{B}_i)} \mathbf{C}^{(\mathcal{B}_i)} \boldsymbol{\varepsilon}^{(\mathcal{B}_i)}, \quad \mathbf{C}^{(\mathcal{B}_i)\dagger} \mathbf{S}^{(\mathcal{B}_i)} \mathbf{C}^{(\mathcal{B}_i)} = \mathbf{1}. \tag{S45}$$

The $j$-th columns of $\mathbf{C}^{(\mathcal{B}_i)}$ represent orthonormal eigenstates $\psi_j$ in $\mathcal{H}_i$ (i.e. each $\mathbf{C}_j^{(\mathcal{B}_i)}$ satisfy the second condition above).

Our goal is to construct a reduced representation of $\mathbf{M}^{(\mathcal{B}_1)}$ in $\mathcal{H}_2$ that projects a selection of eigenstates and eigenvalues, while preserving equivariance. We obtain the projected representation of the eigenstates $\mathbf{C}^{(\mathcal{B}_1)}$ in $\mathcal{B}_2$, $\mathbf{C}^{(\mathcal{B}_1 \to \mathcal{B}_2)}$ as

$$\mathbf{C}^{(\mathcal{B}_1 \to \mathcal{B}_2)} = S^{(\mathcal{B}_2\mathcal{B}_1)} \mathbf{C}^{(\mathcal{B}_1)}.$$

(When only a subset of columns is retained, $\overline{\mathbf{C}}^{(\mathcal{B}_1 \to \mathcal{B}_2)}$ and the corresponding eigenvalues refer to that subset.)

However, the columns of $\mathbf{C}^{(\mathcal{B}_1 \to \mathcal{B}_2)}$ generally do not satisfy the orthonormality condition in Eq. (S45), with respect to $\mathbf{S}^{(\mathcal{B}_2)}$.

$$\left(\mathbf{C}^{(\mathcal{B}_1 \to \mathcal{B}_2)\dagger} \mathbf{S}^{(\mathcal{B}_2)} \mathbf{C}^{(\mathcal{B}_1 \to \mathcal{B}_2)}\right) = \mathbf{C}^{(\mathcal{B}_1)\dagger} S^{(\mathcal{B}_2\mathcal{B}_1)\dagger} \mathbf{S}^{(\mathcal{B}_2)} S^{(\mathcal{B}_2\mathcal{B}_1)} \mathbf{C}^{(\mathcal{B}_1)}. \tag{S46}$$

14

Even using (S43), the RHS does not evaluate to $\mathbf{1}$. One can Löwdin orthogonalize the RHS explicitly by defining the quantity in Eq. (S46) as $\mathbf{G}$, so that, $\mathbf{G} = \mathbf{C}^{(\mathcal{B}_1)\dagger} S^{(\mathcal{B}_2\mathcal{B}_1)\dagger}\mathbf{S}^{(\mathcal{B}_2)} S^{(\mathcal{B}_2\mathcal{B}_1)}\mathbf{C}^{(\mathcal{B}_1)}$ followed by orthogonalizing the projection as $\overline{\mathbf{C}}^{(\mathcal{B}_1 \to \mathcal{B}_2)} = \mathbf{C}^{(\mathcal{B}_1 \to \mathcal{B}_2)} \mathbf{G}^{-1/2}$.

The orthogonalized projection satisfies, $\overline{\mathbf{C}}^{(\mathcal{B}_1 \to \mathcal{B}_2)\dagger} \mathbf{S}^{(\mathcal{B}_2)} \overline{\mathbf{C}}^{(\mathcal{B}_1 \to \mathcal{B}_2)} = \mathbf{1}$, which can be proved easily by substituting in (S46). Note that the orthogonalization requires $\mathbf{G}$ to be invertible, as well as be positive definite. We can show that $\mathbf{G}$ is positive semi-definite as follows. For any column vector $\mathbf{c} \in \mathbf{C}^{(\mathcal{B}_1)}$,

$$\mathbf{c}^\dagger \mathbf{G} \mathbf{c} = \mathbf{c}^\dagger \mathbf{C}^{(\mathcal{B}_1 \to \mathcal{B}_2)\dagger} \mathbf{S}^{(\mathcal{B}_2)} \mathbf{C}^{(\mathcal{B}_1 \to \mathcal{B}_2)} \mathbf{c}$$

Defining $\mathbf{d} = \mathbf{C}^{(\mathcal{B}_1 \to \mathcal{B}_2)} \mathbf{c}$. Then $\mathbf{c}^\dagger \mathbf{G} \mathbf{c} = \mathbf{d}^\dagger \mathbf{S}^{(\mathcal{B}_2)} \mathbf{d}$.

Since $\mathbf{S}^{(\mathcal{B}_2)}$ is the AO overlap matrix, it is Hermitian and positive definite on $\mathcal{H}_2$, implying $\mathbf{d}^\dagger \mathbf{S}^{(\mathcal{B}_2)} \mathbf{d} \geq 0$, with equality iff $\mathbf{d} = 0$. Therefore, $\mathbf{G}$ is Hermitian and positive semi-definite.

A zero eigenvalue of $\mathbf{G}$ corresponds to a nonzero vector $\mathbf{c}$ such that $\mathbf{G}\mathbf{c} = 0$. Then $0 = \mathbf{c}^\dagger \mathbf{G} \mathbf{c} = \mathbf{d}^\dagger \mathbf{S}^{(\mathcal{B}_2)} \mathbf{d}$. The positive definiteness of $\mathbf{S}^{(\mathcal{B}_2)}$ implies $\mathbf{d} = \mathbf{0}$, or equivalently $\mathbf{C}^{(\mathcal{B}_1 \to \mathcal{B}_2)} \mathbf{c} = 0$. That is, zero eigenvalues occur when the columns of $\mathbf{C}^{(\mathcal{B}_1 \to \mathcal{B}_2)}$ become linearly dependent (or vanish) in the reduced basis $\mathcal{B}_2$. Since we require $\mathbf{G}$ to be invertible to perform the Löwdin orthogonalization, we restrict the subspace of projected $\mathbf{C}^{(\mathcal{B}_1)}$ so that $\mathbf{G}$ is strictly positive definite. A subspace spanned by columns of $\mathbf{C}^{(B_1)}$ is representable in $\mathcal{H}_2$ iff $\mathbf{G} > 0$ on that subspace.

In practice, one diagonalizes $\mathbf{G} = \mathbf{U} \operatorname{diag}(\lambda_i) \mathbf{U}^\dagger$ and retains only eigenvectors with $\lambda_i > \epsilon$ ($\epsilon > 0$ is a small threshold), defining the maximal representable subspace. The eigenvectors of $\mathbf{G}$ correspond to linear combinations of the selected eigenstates in $\mathcal{B}_1$.

Although only eigenstates whose projections correspond to $\lambda_i > \varepsilon$ are representable in $\mathcal{B}_2$, it is not known a priori which eigenstates in $\mathcal{B}_1$ satisfy this condition. Representability depends on the overlap between $\mathcal{B}_1$ and $\mathcal{B}_2$, as well as the constraints from symmetry, for instance certain symmetry irreps in one basis can not map onto the other. We therefore begin with an initial subset of eigenstates in $\mathcal{B}_1$, and adaptively expand this candidate state until we obtain a set of $N_2$ eigenstates yielding strictly positive eigenvalues of $\mathbf{G}$.

Having obtained orthonormal projected coefficients $\tilde{\mathbf{C}}^{(\mathcal{B}_2)}$, the reduced operator is

$$\mathbf{M}^{(\mathcal{B}_1 \to \mathcal{B}_2)} = \overline{\mathbf{C}}^{(\mathcal{B}_1 \to \mathcal{B}_2)\dagger} \operatorname{diag}(\varepsilon^{(\mathcal{B}_1)}) \overline{\mathbf{C}}^{(\mathcal{B}_1 \to \mathcal{B}_2)},$$

where $\boldsymbol{\varepsilon}^{(\mathcal{B}_1)}$ is subset of eigenvalues of $\mathbf{M}^{(\mathcal{B}_1)}$ that are projected.

For *effective Op2Op* models, we first obtain the effective reduced operator representation $\mathbf{M}^{(\mathcal{B}_1 \to \mathcal{B}_2)}$ following the procedure described above. This initial projection is subsequently refined through the loss on observables computed through Eq. (17).

[1] J. Nigam, M. J. Willatt, and M. Ceriotti, Equivariant Representations for Molecular Hamiltonians and $N$-Center Atomic-Scale Properties, J. Chem. Phys. **156**, 014 (2022).

[2] J. Nigam, S. Pozdnyakov, and M. Ceriotti, Recursive evaluation and iterative contraction of $N$-body equivariant features, J. Chem. Phys. **153**, 121101 (2020).

[3] M. J. Willatt, F. Musil, and M. Ceriotti, Atom-density representations for machine learning, J. Chem. Phys. **150**, 154110 (2019).

[4] P. W. Battaglia, J. B. Hamrick, V. Bapst, A. Sanchez-Gonzalez, V. Zambaldi, M. Malinowski, A. Tacchetti, D. Raposo, A. Santoro, R. Faulkner, *et al.*, Relational inductive biases, deep learning, and graph networks, arXiv preprint arXiv:1806.01261 (2018).

[5] J. Gilmer, S. S. Schoenholz, P. F. Riley, O. Vinyals, and G. E. Dahl, Neural message passing for quantum chemistry, in *Int. Conf. Mach. Learn.* (2017) pp. 1263–1272.

[6] Q. Jiang, C. Wang, M. Lones, and W. Pang, Demystifying mpnns: Message passing as merely efficient matrix multiplication, arXiv preprint arXiv:2502.00140 (2025).

[7] C. Godsil and G. Royle, *Algebraic Graph Theory (Graduate Texts in Mathematics 207) ISBN 0-387-95220-9 (Springer, New York, 2001).*, Vol. 34 (Cambridge University Press, 2002) pp. 619–631.

[8] L. C. Biedenharn and J. D. Louck, *The Racah-Wigner algebra in Quantum Theory, Encyclopedia of Mathematics and its Applications* (Addison-Wesley, 1981).

[9] S. F. Boys, Electronic wave functions-i. a general method of calculation for the stationary states of any molecular system, Proceedings of the Royal Society of London. Series A. Mathematical and Physical Sciences **200**, 542 (1950).

[10] S. Obara and A. Saika, Efficient recursive computation of molecular integrals over cartesian gaussian functions, The Journal of chemical physics **84**, 3963 (1986).

[11] Z. Pu and Q. Sun, Enhancing pyscf-based quantum chemistry simulations with modern hardware, algorithms, and python tools, APL Computational Physics **1**, 016101 (2025).



\end{document}

%% file: defs.tex
\newcommand{\mcal}[1]{\ensuremath{\mathcal{#1}}}
\newcommand{\mbf}[1]{\ensuremath{\mathbf{#1}}}
\newcommand{\vext}{\mbf{V}}

\NewDocumentCommand{\rep}{s d<| d|>}{%
\IfBooleanTF{#1}{
   \IfValueTF{#2}{
       \IfValueTF{#3}{\braket{#2}{#3}}{\bra{#2}}
       }{
       \IfValueTF{#3}{\ket{#3}}{}
       }
   }{
   \IfValueTF{#2}{
       \IfValueTF{#3}{\braket*{#2}{#3}}{\bra*{#2}}
       }{
       \IfValueTF{#3}{\ket*{#3}}{}
       }
   }
}

\NewDocumentCommand{\rbra}{sm}{\IfBooleanTF{#1}{\rep*<#2|}{\rep<#2|}}
\NewDocumentCommand{\rket}{sm}{\IfBooleanTF{#1}{\rep*|#2>}{\rep|#2>}}
\NewDocumentCommand{\rbraket}{smom}{
    \IfBooleanTF{#1}{
        \IfNoValueTF{#3}{\rep*<#2||#4>}{\rep*<#2|#3\rep*|#4>}
    }{
        \IfNoValueTF{#3}{\rep<#2||#4>}{\rep<#2|#3\rep|#4>}
    }
}

\NewDocumentCommand{\cg}{m m m}{\rep<#1; #2||#3>}
\NewDocumentCommand{\field}{o m e{_} e{^} o e{_} e{^}}{
\IfValueTF{#5}{
  #2\IfValueT{#3}{_#3}\IfValueT{#4}{^{\otimes #4}} 
  \otimes
  #5\IfValueT{#6}{_#6}\IfValueT{#7}{^{\otimes #7}} 
  \IfValueT{#1}{;#1}
}{
  \IfValueTF{#4}{
     #2\IfValueT{#3}{_#3}\IfValueT{#4}{^{\otimes #4}}
     \IfValueT{#1}{;#1}
  }
  {#2\IfValueT{#3}{_#3}}
}
}
\NewDocumentCommand{\frho}{o e{_} e{^}}{
\field[#1]{\rho}_{#2}^{#3}
}

\newcommand{\br}{\mbf{r}}

\newcommand{\bx}{\mbf{x}}

\newcommand{\bR}{\mbf{R}}

\newcommand{\e}{a}  

\NewDocumentCommand{\ex}{e_}{
\IfValueTF{#1}{\e_{#1}\bx_{#1}}{\e\bx}
}  

\NewDocumentCommand\te{s}{\tilde{\e}\IfBooleanTF{#1}{'}{}}
\NewDocumentCommand\tn{s}{\tilde{n}\IfBooleanTF{#1}{'}{}}
\NewDocumentCommand\tl{s}{\tilde{l}\IfBooleanTF{#1}{'}{}}
\NewDocumentCommand\tm{s}{\tilde{m}\IfBooleanTF{#1}{'}{}}
\NewDocumentCommand\tlm{s}{\IfBooleanTF{#1}{\tl*\tm*}{\tl\tm}}
\NewDocumentCommand\tnlm{s}{\IfBooleanTF{#1}{\tnl*\tm*}{\tnl\tm}}
\NewDocumentCommand\tnl{s}{\IfBooleanTF{#1}{\tn*\tl*}{\tn\tl}}

\NewDocumentCommand{\lm}{e_}{
\IfValueTF{#1}{l_{#1}m_{#1}}{lm}
}
\NewDocumentCommand{\nlm}{e_}{
\IfValueTF{#1}{n_{#1}\lm_{#1}}{n\lm}
}
\NewDocumentCommand{\enlm}{e_}{
\IfValueTF{#1}{\e_{#1}\nlm_{#1}}{\e\nlm}
}
\NewDocumentCommand{\en}{e_}{
\IfValueTF{#1}{\e_{#1}n_{#1}}{\e n}
}
\NewDocumentCommand{\nlk}{e_}{
\IfValueTF{#1}{n_{#1}l_{#1}k_{#1}}{nlk}
}
\NewDocumentCommand{\enlk}{e_}{
\IfValueTF{#1}{\e_{#1}\nlk_{#1}}{\e\nlk}
}
\NewDocumentCommand{\enl}{e_}{
\IfValueTF{#1}{\en_{#1}l_#1}{\en l}
}
\NewDocumentCommand{\nl}{e_}{
\IfValueTF{#1}{n_{#1}l_#1}{n l}
}
\NewDocumentCommand{\nnl}{s}{
\IfBooleanTF{#1}{n_1 n_2 l}{n_1; n_2; l}
}
\NewDocumentCommand{\ennl}{s}{
\IfBooleanTF{#1}{\en_1 \en_2 l}{\en_1; \en_2; l}
}

\NewDocumentCommand{\gslm}{s}{
\IfBooleanTF{#1}{\sigma\lambda\mu}{\sigma;\lambda\mu}
}
\NewDocumentCommand{\glm}{}{\lambda\mu}

\newcommand{\Rhat}{\hat{R}}

\NewDocumentCommand{\plm}{s e_}{
  \IfBooleanTF{#1}{l' m'}{
    \IfValueTF{#2}{l_{#2} m_{#2}}{l m}
  }
}

\NewDocumentCommand{\pnlm}{s e_}{
  \IfBooleanTF{#1}{n' l' m'}{
    \IfValueTF{#2}{n_{#2} l_{#2} m_{#2}}{n l m}
  }
}

\NewDocumentCommand{\pnl}{s e_}{
  \IfBooleanTF{#1}{n' l'}{
    \IfValueTF{#2}{n_{#2} l_{#2}}{n l}
  }
}

%% file: refs.bib
@article{batatia2023equivariant,
  title={Equivariant matrix function neural networks},
  author={Batatia, Ilyes and Schaaf, Lars L and Chen, Huajie and Cs{\'a}nyi, G{\'a}bor and Ortner, Christoph and Faber, Felix A},
  journal={arXiv preprint arXiv:2310.10434},
  year={2023}
}

@article{niga+22jcp,
  title = {{Equivariant Representations for Molecular {{Hamiltonians}} and {{{\emph{N}}}} -Center Atomic-Scale Properties}},
  author = {Nigam, Jigyasa and Willatt, Michael J. and Ceriotti, Michele},
  year = {2022},
  journal = {J. Chem. Phys.},
  volume = {156},
  number = {1},
  pages = {014--115},
}

@article{suman2025exploring,
  title={{Exploring the Design Space of Machine Learning Models for Quantum Chemistry with a Fully Differentiable Framework}},
  author={Suman, Divya and Nigam, Jigyasa and Saade, Sandra and Pegolo, Paolo and T\"urk, Hanna and Zhang, Xing and Chan, Garnet Kin-Lic and Ceriotti, Michele},
  journal={Journal of Chemical Theory and Computation},
  year={2025},
}

@article{cignoni2024electronic,
  title={Electronic excited states from physically constrained machine learning},
  author={Cignoni, Edoardo and Suman, Divya and Nigam, Jigyasa and Cupellini, Lorenzo and Mennucci, Benedetta and Ceriotti, Michele},
  journal={ACS Central Science},
  volume={10},
  number={3},
  pages={637--648},
  year={2024},
  publisher={ACS Publications}
}

@article{yin2024harmonizingdeephe3,
  title={{Harmonizing Covariance and Expressiveness for Deep Hamiltonian Regression in Crystalline Material Research: a Hybrid Cascaded Regression Framework}},
  author={Yin, Shi and Zhu, Xudong and Gao, Tianyu and Zhang, Haochong and Wu, Feng and He, Lixin},
  journal={arXiv:2401.00744},
  year={2024}
}

@article{zhong2024universal,
  title={{Universal Machine Learning Kohn-Sham Hamiltonian for Materials}},
  author={Zhong, Yang and Yang, Jihui and Xiang, Hongjun and Gong, Xingao},
  journal={arXiv:2402.09251},
  year={2024}
}

@article{sun2018pyscf,
  title={{PySCF: the Python-based Simulations of Chemistry Framework}},
  author={Sun, Qiming and Berkelbach, Timothy C and Blunt, Nick S and Booth, George H and Guo, Sheng and Li, Zhendong and Liu, Junzi and McClain, James D and Sayfutyarova, Elvira R and Sharma, Sandeep and others},
  journal={Wiley Interdisciplinary Reviews: Computational Molecular Science},
  volume={8},
  number={1},
  pages={e1340},
  year={2018},
}

@article{sun2022molecular,
    author = {Sun, Jiace and Cheng, Lixue and Miller, Thomas F., III},
    title = {{Molecular Dipole Moment Learning via Rotationally Equivariant Derivative Kernels in Molecular-orbital-based Machine Learning}},
    journal = {The Journal of Chemical Physics},
    volume = {157},
    number = {10},
    pages = {104--109},
    year = {2022},
}

@article{montavon2013machine,
  title={{Machine Learning of Molecular Electronic Properties in Chemical Compound Space}},
  author={Montavon, Gr{\'e}goire and Rupp, Matthias and Gobre, Vivekanand and Vazquez-Mayagoitia, Alvaro and Hansen, Katja and Tkatchenko, Alexandre and M{\"u}ller, Klaus-Robert and Von Lilienfeld, O Anatole},
  journal={New Journal of Physics},
  volume={15},
  number={9},
  pages={095--003},
  year={2013},
}

@article{febrer2024graph2mat,
  title={{Graph2Mat: Universal Graph to Matrix Conversion for Electron Density Prediction}},
  author={Febrer, Pol and J{\o}rgensen, Peter and Pruneda, Miguel and Garcia, Alberto and Ordejon, Pablo and Bhowmik, Arghya},
  year={2024}
}

@article{hegde2017machine,
  title={{Machine-learned Approximations to Density Functional Theory Hamiltonians}},
  author={Hegde, Ganesh and Bowen, R Chris},
  journal={Scientific reports},
  volume={7},
  number={1},
  pages={1--11},
  year={2017},
}

@article{schu+19nc,
  title = {{Unifying Machine Learning and Quantum Chemistry with a Deep Neural Network for Molecular Wavefunctions}},
  author = {Sch{\"u}tt, Kristoff T. and Gastegger, Michael and Tkatchenko, Alexandre and M{\"u}ller, Klaus-Robert and Maurer, Reinhard J.},
  year = {2019},
  journal = {Nat Commun},
  volume = {10},
  number = {1},
  pages = {50--24},
}

@article{west-maur21cs,
  title = {{Physically Inspired Deep Learning of Molecular Excitations and Photoemission Spectra}},
  author = {Westermayr, Julia and Maurer, Reinhard J.},
  year = {2021},
  journal = {Chem. Sci.},
  volume = {12},
  number = {32},
  pages = {10755--10764},
}

@article{unke2021se3equivariant,
title={{SE(3)-equivariant Prediction of Molecular Wavefunctions and Electronic Densities}}, 
author={Oliver T. Unke and Mihail Bogojeski and Michael Gastegger and Mario Geiger and Tess Smidt and Klaus-Robert Müller},
journal={NeurIPS},
number={1106},
pages={14434--14447},
year={2021}
}

@article{grisafi2018transferable,
  title={Transferable Machine-Learning Model of the Electron Density},
  author={Grisafi, Andrea and Fabrizio, Alberto and Meyer, Benjamin and Wilkins, David M and Corminboeuf, Clemence and Ceriotti, Michele},
  journal={ACS Central Science},
  volume={5},
  number={1},
  pages={57--64},
  year={2018},
}

@article{qiao2020orbnet,
  title={{OrbNet: Deep Learning for Quantum Chemistry Using Symmetry-adapted Atomic-orbital Features}},
  author={Qiao, Zhuoran and Welborn, Matthew and Anandkumar, Animashree and Manby, Frederick R and Miller III, Thomas F},
  journal={The Journal of Chemical Physics},
  volume={153},
  number={12},
  pages={124--111},
  year={2020},
}

@article{welborn2018transferability,
  title={{Transferability in Machine Learning for Electronic Structure via the Molecular Orbital Basis}},
  author={Welborn, Matthew and Cheng, Lixue and Miller III, Thomas F},
  journal={Journal of Chemical Theory and Computation},
  volume={14},
  number={9},
  pages={4772--4779},
  year={2018},
}

@article{fabrizio2021spa,
  title={{{SPA$\hat{H}$M}: The Spectrum of Approximated Hamiltonian Matrices representations}},
  author={Fabrizio, Alberto and Briling, Ksenia R and Corminboeuf, Clemence},
  journal={Digital Discovery},
  number={1},
  pages={286--294},
  year={2022}
}

@article{rackers2023recipe,
  title={{A Recipe for Cracking the Quantum Scaling Limit with Machine Learned Electron Densities}},
  author={Rackers, Joshua A and Tecot, Lucas and Geiger, Mario and Smidt, Tess E},
  journal={Machine Learning: Science and Technology},
  volume={4},
  number={1},
  pages={015--027},
  year={2023},
}

@article{shao2023machine,
  title={Machine learning electronic structure methods based on the one-electron reduced density matrix},
  author={Shao, Xuecheng and Paetow, Lukas and Tuckerman, Mark E and Pavanello, Michele},
  journal={Nature communications},
  volume={14},
  number={1},
  pages={6281},
  year={2023},
  publisher={Nature Publishing Group UK London}
}

@article{koker2024higher,
  title={Higher-order Equivariant Neural Networks for Charge Density Prediction in Materials},
  author={Koker, Teddy and Quigley, Keegan and Taw, Eric and Tibbetts, Kevin and Li, Lin},
  journal={npj Computational Materials},
  volume={10},
  number={1},
  pages={161},
  year={2024},
}

@article{zhang2022equivariant,
  title={{Equivariant Analytical Mapping of First Principles Hamiltonians to Accurate and Transferable Materials Models}},
  author={Zhang, Liwei and Onat, Berk and Dusson, Genevi{\`e}ve and McSloy, Adam and Anand, Gautam and Maurer, Reinhard J and Ortner, Christoph and Kermode, James R},
  journal={npj Computational Materials},
  volume={8},
  number={1},
  pages={158},
  year={2022},
}

@article{kirkpatrick2021pushing,
  title={Pushing the Frontiers of Density Functionals by Solving the Fractional Electron Problem},
  author={Kirkpatrick, James and McMorrow, Brendan and Turban, David HP and Gaunt, Alexander L and Spencer, James S and Matthews, Alexander GDG and Obika, Annette and Thiry, Louis and Fortunato, Meire and Pfau, David and others},
  journal={Science},
  volume={374},
  number={6573},
  pages={1385--1389},
  year={2021},
}

@article{niga+20jcp,
  title = {Recursive Evaluation and Iterative Contraction of {{{\emph{N}}}} -Body Equivariant Features},
  author = {Nigam, Jigyasa and Pozdnyakov, Sergey and Ceriotti, Michele},
  year = {2020},
  month = sep,
  journal = {J. Chem. Phys.},
  volume = {153},
  number = {12},
  pages = {121101},
  issn = {0021-9606, 1089-7690},
  doi = {10.1063/5.0021116},
  langid = {english}
}

@article{veit+20jcp,
  title = {{Predicting Molecular Dipole Moments by Combining Atomic Partial Charges and Atomic Dipoles}},
  author = {Veit, Max and Wilkins, David M. and Yang, Yang and DiStasio, Robert A. and Ceriotti, Michele},
  year = {2020},
  journal = {J. Chem. Phys.},
  volume = {153},
  number = {2},
  pages = {024--113},
}

@article{li2024image,
  title={{Image Super-resolution Inspired Electron Density Prediction}},
  author={Li, Chenghan and Sharir, Or and Yuan, Shunyue and Chan, Garnet K},
  journal={arXiv:2402.12335},
  year={2024}
}

@article{haldar2025gears,
  title={{GEARS H: Accurate Machine-learned Hamiltonians for Next-Generation Device-scale Modeling}},
  author={Haldar, Anubhab and Hamze, Ali K and Sivadas, Nikhil and Shin, Yongwoo},
  journal={arXiv:2506.10298},
  year={2025}
}

@article{tang2024approaching,
  title={Approaching coupled-cluster accuracy for molecular electronic structures with multi-task learning},
  author={Tang, Hao and Xiao, Brian and He, Wenhao and Subasic, Pero and Harutyunyan, Avetik R and Wang, Yao and Liu, Fang and Xu, Haowei and Li, Ju},
  journal={Nature Computational Science},
  pages={1--11},
  year={2024},
  publisher={Nature Publishing Group US New York}
}

@article{luise2025skala,
	title = {Accurate and Scalable Exchange-correlation with Deep Learning},
	author = {Luise, Giulia and Huang, Chin-Wei and Vogels, Thijs and Kooi, Derk P and Ehlert, Sebastian and Lanius, Stephanie and Giesbertz, Klaas J H and Karton, Amir and Gunceler, Deniz and Stanley, Megan and Bruinsma, Wessel P and Huang, Lin and Wei, Xinran and Torres, José Garrido and Katbashev, Abylay and Máté, Bálint and Kaba, Sékou-Oumar and Sordillo, Roberto and Chen, Yingrong and Williams-Young, David B and Bishop, Christopher M and Hermann, Jan},
    journal={arXiv:2506.14665},
year = {2025},
}

@article{nequip,
    author = {Batzner, S. and Musaelian, A. and Sun, L. and Geiger, M. and Mailoa, J. P. and Kornbluth, M. and Molinari, N. and Schmidt, T. and Kozinsky, B.},
    title = {{E(3)-Equivariant Graph Neural Networks for Data-efficient and Accurate Interatomic Potentials}},
    journal = {Nature Communications},
    volume = {13},
    number = {2453},
    year = {2022}
}

@inproceedings{mace,
 author = {Batatia, Ilyes and Kovacs, David P and Simm, Gregor and Ortner, Christoph and Csanyi, Gabor},
 booktitle = {NeurIPS},
 editor = {S. Koyejo and S. Mohamed and A. Agarwal and D. Belgrave and K. Cho and A. Oh},
 pages = {11423--11436},
 publisher = {Curran Associates, Inc.},
 title = {{MACE: Higher Order Equivariant Message Passing Neural Networks for Fast and Accurate Force Fields}},
 volume = {35},
 year = {2022}
}

@article{musaelian2023learning,
  title={{Learning Local Equivariant Representations for Large-Scale Atomistic Dynamics}},
  author={Musaelian, Albert and Batzner, Simon and Johansson, Anders and Sun, Lixin and Owen, Cameron J and Kornbluth, Mordechai and Kozinsky, Boris},
  journal={Nature Communications},
  volume={14},
  number={1},
  pages={579},
  year={2023},
}

@article{behler2021four,
  title={Four Generations of High-dimensional Neural Network Potentials},
  author={Behler, Jorg},
  journal={Chemical Reviews},
  volume={121},
  number={16},
  pages={10037--10072},
  year={2021},
  publisher={ACS Publications}
}

@article{behl-parr07prl,
author = {Behler, J{\"{o}}rg and Parrinello, Michele},



journal = {Phys. Rev. Lett.},


pages = {146401},
title = {{Generalized Neural-Network Representation of High-Dimensional Potential-Energy Surfaces}},

volume = {98},
year = {2007}
}

@article{gap2013,
  title = {{Gaussian Approximation Potentials: The Accuracy of Quantum Mechanics, without the Electrons}},
  author = {Bart\'ok, Albert P. and Payne, Mike C. and Kondor, Risi and Cs\'anyi, G\'abor},
  journal = {Phys. Rev. Lett.},
  volume = {104},
  issue = {13},
  pages = {136--403},
  numpages = {4},
  year = {2010}
}

@article{rupp+12prl,
author = {Rupp, Matthias and Tkatchenko, Alexandre and M{\"{u}}ller, Klaus-Robert and von Lilienfeld, O. Anatole},
journal = {Phys. Rev. Lett.},
pages = {058301},
title = {{Fast and Accurate Modeling of Molecular Atomization Energies with Machine Learning}},
volume = {108},
year = {2012}
}

@article{shap16mms,
author = {Shapeev, Alexander V.},
journal = {Multiscale Modeling {\&} Simulation},
pages = {1153--1173},
title = {{Moment Tensor Potentials: A Class of Systematically Improvable Interatomic Potentials}},
volume = {14},
year = {2016}
}

@article{drau19prb,
author = {Drautz, Ralf},
journal = {Phys. Rev. B},
pages = {014104},
publisher = {American Physical Society},
title = {{Atomic Cluster Expansion for Accurate and Transferable Interatomic Potentials}},
volume = {99},
year = {2019}
}

@article{geiger2022e3nn,
  title={e3nn: Euclidean neural networks},
  author={Geiger, Mario and Smidt, Tess},
  journal={arXiv preprint arXiv:2207.09453},
  year={2022}
}

@article{bigi+22jcp,
  title = {A Smooth Basis for Atomistic Machine Learning},
  author = {Bigi, Filippo and {Huguenin-Dumittan}, Kevin K. and Ceriotti, Michele and Manolopoulos, David E.},
  year = {2022},
  month = dec,
  journal = {J. Chem. Phys.},
  volume = {157},
  number = {23},
  pages = {234101},
  issn = {0021-9606, 1089-7690},
  doi = {10.1063/5.0124363},
  urldate = {2022-12-17},
  langid = {english}
}

@article{hohenberg1964inhomogeneous,
  title={Inhomogeneous electron gas},
  author={Hohenberg, Pierre and Kohn, Walter},
  journal={Physical review},
  volume={136},
  number={3B},
  pages={B864},
  year={1964},
  publisher={APS}
}

@article{brockherde2017bypassing,
  title={Bypassing the Kohn-Sham equations with machine learning},
  author={Brockherde, Felix and Vogt, Leslie and Li, Li and Tuckerman, Mark E and Burke, Kieron and M{\"u}ller, Klaus-Robert},
  journal={Nature communications},
  volume={8},
  number={1},
  pages={872},
  year={2017},
  publisher={Nature Publishing Group UK London}
}

@article{rana2025learning,
  title={Learning the One-Electron Reduced Density Matrix at SCF Convergence Thresholds},
  author={Rana, Bhaskar and Viot, Nicolas and Martinez B, Jessica A and Shao, Xuecheng and Ramos, Pablo and Pavanello, Michele},
  journal={Journal of Chemical Theory and Computation},
  year={2025},
  publisher={ACS Publications}
}

@article{will+19jcp,
  title = {Atom-Density Representations for Machine Learning},
  author = {Willatt, Michael J. and Musil, F{\'e}lix and Ceriotti, Michele},
  year = {2019},
  month = apr,
  journal = {J. Chem. Phys.},
  volume = {150},
  number = {15},
  pages = {154110},
  issn = {00219606},
  doi = {10.1063/1.5090481}
}

@article{bart+13prb,
  title = {On Representing Chemical Environments},
  author = {Bart{\'o}k, Albert P. and Kondor, Risi and Cs{\'a}nyi, G{\'a}bor},
  year = {2013},
  month = may,
  journal = {Phys. Rev. B},
  volume = {87},
  number = {18},
  pages = {184115},
  issn = {1098-0121},
  doi = {10.1103/PhysRevB.87.184115},
  urldate = {2014-09-24}
}

@article{niga+22jcp2,
  title = {Unified Theory of Atom-Centered Representations and Message-Passing Machine-Learning Schemes},
  author = {Nigam, Jigyasa and Pozdnyakov, Sergey and Fraux, Guillaume and Ceriotti, Michele},
  year = {2022},
  month = may,
  journal = {J. Chem. Phys.},
  volume = {156},
  number = {20},
  pages = {204115},
  issn = {0021-9606, 1089-7690},
  doi = {10.1063/5.0087042},
  urldate = {2022-05-28},
  langid = {english}
}

@article{gris+18prl,
  title = {Symmetry-{{Adapted Machine Learning}} for {{Tensorial Properties}} of {{Atomistic Systems}}},
  author = {Grisafi, Andrea and Wilkins, David M. and Cs{\'a}nyi, G{\'a}bor and Ceriotti, Michele},
  year = {2018},
  month = jan,
  journal = {Phys. Rev. Lett.},
  volume = {120},
  number = {3},
  pages = {036002},
  issn = {10797114},
  doi = {10.1103/PhysRevLett.120.036002}
}

@misc{matcloud18c,
  title = {Dataset: Symmetry-{{Adapted Machine Learning}} for {{Tensorial Properties}} of {{Atomistic Systems}}},
  author = {Grisafi, Andrea and Wilkins, David M. and Cs{\'a}nyi, Gabor and Ceriotti, Michele},
  year = {2018},
  doi = {10.24435/materialscloud:2018.0009/v1},
  howpublished = {\url{http://doi.org/10.24435/materialscloud:2018.0009/v1}}
}

@article{dusson2022atomic,
  title={Atomic cluster expansion: Completeness, efficiency and stability},
  author={Dusson, Genevieve and Bachmayr, Markus and Cs{\'a}nyi, G{\'a}bor and Drautz, Ralf and Etter, Simon and van Der Oord, Cas and Ortner, Christoph},
  journal={Journal of Computational Physics},
  volume={454},
  pages={110946},
  year={2022},
  publisher={Elsevier}
}

@article{kipf2016semi,
  title={Semi-supervised classification with graph convolutional networks},
  author={Kipf, TN and Welling, Max},
  journal={arXiv preprint arXiv:1609.02907},
  year={2016}
}

@article{pozdnyakov2020completeness,
  title={On the Completeness of Atomic Structure Representations},
  author={Pozdnyakov, Sergey N and Willatt, Michael J and Bart{\'o}k, Albert P and Ortner, Christoph and Cs{\'a}nyi, G{\'a}bor and Ceriotti, Michele},
  journal={arXiv preprint arXiv:2001.11696},
  year={2020}
}

@article{goscinski2021role,
  title={The role of feature space in atomistic learning},
  author={Goscinski, Alexander and Fraux, Guillaume and Imbalzano, Giulio and Ceriotti, Michele},
  journal={Machine Learning: Science and Technology},
  volume={2},
  number={2},
  pages={025028},
  year={2021},
  publisher={IOP Publishing}
}

@article{goscinski2021optimal,
  title={Optimal radial basis for density-based atomic representations},
  author={Goscinski, Alexander and Musil, F{\'e}lix and Pozdnyakov, Sergey and Nigam, Jigyasa and Ceriotti, Michele},
  journal={The Journal of Chemical Physics},
  volume={155},
  number={10},
  year={2021},
  publisher={AIP Publishing}
}

@article{jiang2025demystifying,
  title={Demystifying MPNNs: Message Passing as Merely Efficient Matrix Multiplication},
  author={Jiang, Qin and Wang, Chengjia and Lones, Michael and Pang, Wei},
  journal={arXiv preprint arXiv:2502.00140},
  year={2025}
}

@article{qian2025equivariant,
  title={Equivariant Electronic Hamiltonian Prediction with Many-Body Message Passing},
  author={Qian, Chen and Vitartas, Valdas and Kermode, James and Maurer, Reinhard J},
  journal={arXiv preprint arXiv:2508.15108},
  year={2025}
}

@article{musi+21cr,
  title = {Physics-{{Inspired Structural Representations}} for {{Molecules}} and {{Materials}}},
  author = {Musil, Felix and Grisafi, Andrea and Bart{\'o}k, Albert P. and Ortner, Christoph and Cs{\'a}nyi, G{\'a}bor and Ceriotti, Michele},
  year = {2021},
  month = aug,
  journal = {Chem. Rev.},
  volume = {121},
  number = {16},
  pages = {9759--9815},
  issn = {0009-2665, 1520-6890},
  doi = {10.1021/acs.chemrev.1c00021},
  urldate = {2021-08-30},
  langid = {english}
}

@article{thom+18arxiv,
  title = {Tensor Field Networks: {{Rotation-}} and Translation-Equivariant Neural Networks for {{3D}} Point Clouds},
  author = {Thomas, Nathaniel and Smidt, Tess and Kearnes, Steven and Yang, Lusann and Li, Li and Kohlhoff, Kai and Riley, Patrick},
  year = {2018},
  journal = {arxiv:1802.08219}
}

@inproceedings{gilm+17icml,
  title = {Neural Message Passing for Quantum Chemistry},
  booktitle = {Int. {{Conf}}. {{Mach}}. {{Learn}}.},
  author = {Gilmer, Justin and Schoenholz, Samuel S and Riley, Patrick F and Vinyals, Oriol and Dahl, George E},
  year = {2017},
  pages = {1263--1272}
}

@article{zhang2022differentiable,
  title={Differentiable quantum chemistry with PySCF for molecules and materials at the mean-field level and beyond},
  author={Zhang, Xing and Chan, Garnet Kin-Lic},
  journal={The Journal of Chemical Physics},
  volume={157},
  number={20},
  pages={204801},
  year={2022},
  publisher={AIP Publishing LLC}
}

@article{choudhary2025slakonet,
  title={Slakonet: A unified slater-koster tight-binding framework using neural network infrastructure for the periodic table},
  author={Choudhary, Kamal},
  journal={The Journal of Physical Chemistry Letters},
  volume={16},
  number={43},
  pages={11109--11119},
  year={2025},
  publisher={ACS Publications}
}

@article{deringer2021gaussian,
  title={Gaussian process regression for materials and molecules},
  author={Deringer, Volker L and Bart{\'o}k, Albert P and Bernstein, Noam and Wilkins, David M and Ceriotti, Michele and Cs{\'a}nyi, G{\'a}bor},
  journal={Chemical reviews},
  volume={121},
  number={16},
  pages={10073--10141},
  year={2021},
  publisher={ACS Publications}
}

@article{zhang2025dm,
      title={{A Symmetry-preserving and Transferable Representation for Learning the Kohn-Sham Density Matrix}}, 
      author={Liwei Zhang and Patrizia Mazzeo and Michele Nottoli and Edoardo Cignoni and Lorenzo Cupellini and Benjamin Stamm},
      year={2025},
      journal={arXiv:2503.08400},
}

@article{darby2022compressing,
  title={Compressing local atomic neighbourhood descriptors},
  author={Darby, James P and Kermode, James R and Cs{\'a}nyi, G{\'a}bor},
  journal={npj Computational Materials},
  volume={8},
  number={1},
  pages={166},
  year={2022},
  publisher={Nature Publishing Group UK London}
}

@article{friede2024dxtb,
  title={dxtb—An efficient and fully differentiable framework for extended tight-binding},
  author={Friede, Marvin and H{\"o}lzer, Christian and Ehlert, Sebastian and Grimme, Stefan},
  journal={The Journal of Chemical Physics},
  volume={161},
  number={6},
  year={2024},
  publisher={AIP Publishing}
}

@article{bai2022machine,
  title={Machine learning the Hohenberg-Kohn map for molecular excited states},
  author={Bai, Yuanming and Vogt-Maranto, Leslie and Tuckerman, Mark E and Glover, William J},
  journal={Nature communications},
  volume={13},
  number={1},
  pages={7044},
  year={2022},
  publisher={Nature Publishing Group UK London}
}

@article{venturella2025unified,
	title = {Unified deep learning framework for many-body quantum chemistry via {Green}’s functions},
	volume = {5},
	copyright = {2025 The Author(s), under exclusive licence to Springer Nature America, Inc.},
	issn = {2662-8457},
	url = {https://www.nature.com/articles/s43588-025-00810-z},
	doi = {10.1038/s43588-025-00810-z},
	number = {6},
	journal = {Nature Computational Science},
	publisher = {Nature Publishing Group},
	author = {Venturella, Christian and Li, Jiachen and Hillenbrand, Christopher and Leyva Peralta, Ximena and Liu, Jessica and Zhu, Tianyu},
	month = jun,
	year = {2025},
	pages = {502--513},
}

@misc{matcloud-anthracene, 
title = {Materials Cloud Dataset Electronic excited states from physically-constrained machine learning}, 
author = {Cignoni, Edoardo and Suman, Divya and Nigam, Jigyasa and Cupellini, Lorenzo and Mennucci, Benedetta and Ceriotti, Michele},
year = {2024},
url={https://doi.org/10.24435/materialscloud:j2-58}}

@article{zhang2026transferable,
	title = {Transferable {Machine} {Learning} of {Electronic} {Hamiltonians} with {Superposition}-of-{Atomic}-{Potentials} {Features}},
	url = {http://arxiv.org/abs/2606.12326},
	doi = {10.48550/arXiv.2606.12326},
	publisher = {arXiv},
	author = {Zhang, Chaoqun and Venturella, Christian and Chen, Enzhi and Zhu, Tianyu},
	month = jun,
	year = {2026},
	note = {arXiv:2606.12326 [physics.chem-ph]},

}
